\documentclass[journal]{IEEEtran}
\IEEEoverridecommandlockouts
\usepackage{booktabs}

\makeatletter
\newsavebox{\@tabnotebox}

\makeatother	
\newif\ifhavebib

\usepackage{blindtext}
\usepackage{graphicx}
\usepackage{graphicx}
\usepackage{array}
\usepackage{url}
\usepackage{algorithmic}
\usepackage[cmex10]{amsmath}
\usepackage{ifpdf}
\usepackage{amssymb,amsmath}
\usepackage{empheq}
\usepackage{stfloats}   
\usepackage{textcomp}
\usepackage{cite}
\usepackage{multirow}
\usepackage{diagbox}
\usepackage{subcaption}
\let\oldFootnote\footnote
\newcommand\nextToken\relax

\renewcommand\footnote[1]{%
    \oldFootnote{#1}\futurelet\nextToken\isFootnote}

\newcommand\isFootnote{%
    \ifx\footnote\nextToken\textsuperscript{,}\fi}

\usepackage[ruled,vlined,lined,ruled,boxed]{algorithm2e}
\usepackage{fixltx2e}
\usepackage[dvipsnames,usenames]{color}
\usepackage[normalem]{ulem}
\usepackage{amsthm}

\usepackage{graphicx}
\usepackage{ifpdf}
\usepackage{cite}
\usepackage{enumerate}
\usepackage{enumitem}

\usepackage{array}
\usepackage{mdwmath}
\usepackage{mdwtab}
\usepackage{eqparbox}
\usepackage{bm}

\usepackage{setspace}
\usepackage{booktabs}
\usepackage[subnum]{cases}
\usepackage{rotating}

\usepackage{listings} 
\definecolor{Red}{rgb}{1,0,0}
\definecolor{Blue}{rgb}{0,0,1}
\definecolor{Olive}{rgb}{0.41,0.55,0.13}
\definecolor{Green}{rgb}{0,1,0}
\definecolor{MGreen}{rgb}{0,0.8,0}
\definecolor{DGreen}{rgb}{0,0.55,0}
\definecolor{Yellow}{rgb}{1,1,0}
\definecolor{Cyan}{rgb}{0,1,1}
\definecolor{Magenta}{rgb}{1,0,1}
\definecolor{Orange}{rgb}{1,.5,0}
\definecolor{Violet}{rgb}{.5,0,.5}
\definecolor{Purple}{rgb}{.75,0,.25}
\definecolor{Brown}{rgb}{.75,.5,.25}
\definecolor{Grey}{rgb}{.5,.5,.5}

\newcommand{\boxhead}[5]{
   \pagestyle{myheadings}
   \thispagestyle{plain}
   \setcounter{page}{1}
   \noindent
   \begin{center}
   \framebox{
      \vbox{\vspace{2mm}
    \hbox to 6.28in { {\bf #1 \hfill} }
       \vspace{6mm}
       \hbox to 6.28in { {\Large \hfill \bf #2  \hfill} }
       \vspace{6mm}
       \hbox to 6.28in { {\it #3 #4 \hfill  #5} }
      \vspace{2mm}}
   }
   \end{center}
   \markboth{#5 -- #2}{#5 -- #2}
   \vspace*{4mm}
}

\usepackage[colorlinks=true, urlcolor=black,  linkcolor=black, citecolor=black]{hyperref}
\theoremstyle{definition}
 
\newtheorem{proposition}{Proposition}

\theoremstyle{remark}
\newtheorem{remark}{Remark}

\theoremstyle{definition}

\DeclarePairedDelimiterX{\infdivx}[2]{(}{)}{%
	#1\;\delimsize\|\;#2%
}

\DeclarePairedDelimiter{\norm}{\lVert}{\rVert}
\DeclarePairedDelimiter{\abs}{\lvert}{\rvert}

\def\tr{\mathop{\rm tr}\nolimits}%
\def\extr{\mathop{\rm extr}\nolimits}%
\def\vect{\mathop{\rm vec}\nolimits}%
\def\diag{\mathop{\rm diag}\nolimits}%
\def\Re{\mathop{\rm Re}\nolimits}%
\newcommand{\av}{{\bf a}}

\newcommand{\Cv}{{\bf C}}
\newcommand{\Bv}{{\bf B}}

\newcommand{\Xv}{{\bf X}}
\newcommand{\Yv}{{\bf Y}}
\newcommand{\Zv}{{\bf Z}}

\newcommand{\Rv}{{\bf R}}
\newcommand{\Qv}{{\bf Q}}

\newcommand{\Av}{{\bf A}}

\newcommand{\Hv}{{\bf H}}
\newcommand{\Gv}{{\bf G}}
\newcommand{\Sv}{{\bf S}}
\newcommand{\Nv}{{\bf N}}
\newcommand{\Iv}{{\bf I}}
\newcommand{\fv}{{\bf f}}
\newcommand{\gv}{{\bf g}}
\newcommand{\xv}{{\bf x}}

\newcommand{\yv}{{\bf y}}

\newcommand{\hv}{{\bf h}}

\newcommand{\nv}{{\bf n}}

\newcommand{\psiv}{\boldsymbol \psi}

\newcommand{\Wv}{{\bf W}}

\newcommand{\varphiv}{\boldsymbol \varphi}

\newcommand{\varthetav}{\boldsymbol \vartheta}
\newcommand{\muv}{\boldsymbol \mu}
\newcommand{\Sigmav}{\boldsymbol \Sigma}

\newcommand{\varsigmav}{\boldsymbol \varsigma}

\newcommand{\Sh}{{\hat{S}}}
\newcommand{\Gh}{{\hat{G}}}

\newcommand{\gh}{{\hat{g}}}

\newcommand{\ph}{{\hat{p}}}
\newcommand{\qh}{{\hat{q}}}
\newcommand{\sh}{{\hat{s}}}

\newcommand{\wh}{{\hat{w}}}

\newcommand{\bh}{{\hat{b}}}
\newcommand{\ch}{{\hat{c}}}
\newcommand{\eh}{{\hat{e}}}

\newcommand{\zh}{{\hat{z}}}
\newcommand{\dih}{{\hat{d}}}

\newcommand{\betah}{\hat \beta}
\newcommand{\gammah}{\hat \gamma}

\newcommand{\alphah}{\hat \alpha}
\newcommand{\muh}{\hat \mu}
\newcommand{\xih}{\hat \xi}
\newcommand{\Hvh}{{\hat{\bf H}}}
\newcommand{\Svh}{{\hat{\bf S}}}
\newcommand{\Gvh}{{\hat{\bf G}}}

\newcommand{\Hvb}{{\bar{\bf H}}}

\newcommand{\Hvt}{{\widetilde{\Hv}}}

\def\a{\alpha}

\def\e{\epsilon}

\DeclareMathOperator\E{E}

 \def\E{\mathbb{E}}

\def\de \mathrm{d}

\newcommand{\Norm}{\mathcal{N}}
\newcommand{\CN}{\mathcal{CN}}

\newcommand\eg{e.g.,\xspace}
\newcommand\ie{i.e.,\xspace}
\def\textiid{i.i.d.\@\xspace}
\newcommand\iid{\ifmmode\text{ i.i.d. } \else \textiid \fi}

\newcommand{\Complex}{\mathbb{C}}
\newcommand{\Real}{\mathbb{R}}

\newcommand{\beqs}{\begin{equation*}}
\newcommand{\eeqs}{\end{equation*}}
\newcommand{\beq}{\begin{equation}}
\newcommand{\eeq}{\end{equation}}
\begin{document}

\setitemize{listparindent=\parindent,partopsep=0pt,topsep=-0.25ex}
\setenumerate{fullwidth,itemindent=\parindent,listparindent=\parindent,itemsep=0ex,partopsep=0pt,parsep=0ex}

\havebibtrue
\title{Matrix-Calibration-Based Cascaded Channel Estimation for Reconfigurable Intelligent Surface Assisted Multiuser MIMO}
\author{
	Hang~Liu,~\IEEEmembership{Graduate Student Member,~IEEE,} Xiaojun~Yuan,~\IEEEmembership{Senior Member,~IEEE,}
	and~Ying-Jun~Angela~Zhang,~\IEEEmembership{Fellow,~IEEE}
	\thanks{The work was supported in part by the General Research Fund (Project number 14208017) established by the Hong Kong Research Grants Council, in part by the National Key R\&D Program of China under grant 2018YFB1801105, and in part by the 111 Project of China under Grant B20064. This work was presented in part at the IEEE International Symposium on Information Theory (ISIT), 2020 \cite{under3}.}%
	\thanks{H. Liu and Y.-J. A. Zhang are with the Department of Information Engineering, The Chinese University of Hong Kong, Shatin, New Territories, Hong Kong SAR (e-mail: lh117@ie.cuhk.edu.hk; yjzhang@ie.cuhk.edu.hk). X. Yuan is with the Center for Intelligent Networking and Communications, the University of Electronic Science and Technology of China, Chengdu, China (e-mail: xjyuan@uestc.edu.cn).}
}
\maketitle
\begin{abstract}
Reconfigurable intelligent surface (RIS) is envisioned to be an essential component of the paradigm for beyond 5G networks as it can potentially provide similar or higher  array gains with much lower hardware cost and energy consumption compared with the massive multiple-input multiple-output (MIMO) technology. In this paper, we focus on one of the fundamental challenges, namely the channel acquisition, in a RIS-assisted multiuser MIMO system. The state-of-the-art channel acquisition approach in such a system with fully passive RIS elements estimates the cascaded transmitter-to-RIS and RIS-to-receiver channels by adopting excessively long training sequences. To estimate the cascaded channels with an affordable training overhead, we formulate the channel estimation problem in the RIS-assisted multiuser MIMO system as a matrix-calibration based matrix factorization task.  By exploiting the information on the slow-varying channel components and the hidden channel sparsity, we propose a novel message-passing based algorithm to factorize the cascaded channels.  Furthermore, we present an analytical framework to characterize the theoretical performance bound of the proposed estimator in the large-system limit. Finally, we conduct simulations to verify the high accuracy and efficiency of the proposed algorithm.
\end{abstract}
\section{Introduction}\label{sec_intro}
As one of the key techniques of the fifth-generation (5G) wireless networks, massive multiple-input multiple-output (MIMO) can greatly enhance system throughput and enlarge cell coverage. However, expensive hardware cost and high power consumption are two unresolved challenges in current massive MIMO systems \cite{MIMO_ES}. Substitute technologies have been explored to achieve more sustainable and more reliable communications for the next generation mobile networks. Among these new technologies, reconfigurable intelligent surface (RIS, a.k.a. passive holographic MIMO surface \cite{LIS_Holographic}) assisted MIMO \cite{LIS_first} is deemed very promising to realize similar or even higher array gains with significant cost reduction compared with massive MIMO. Comprising a large number of reconfigurable reflecting elements that can induce adjustable and independent phase shifts on the incident signals, a RIS is able to constructively combine the reflected  signals to achieve a high level of energy focusing at the receiver side. Due to the passive and low-cost nature of the reflecting elements, the RIS requires very low energy consumption and is easy to be integrated into the existing wireless systems \cite{Metasurface_mag}. 
  
In recent years, the design of RISs in assisting wireless communications has attracted extensive attention. To name a few, the authors in \cite{Metasurface_mag,LIS_array1,LIS_Indoor} employ RISs to control the propagation environment and improve the coverage for indoor communications. In \cite{LIS_Zhangrui,LIS_CHuang,LIS_Assisted,LIS_Yan,LIS_Jinshi}, the authors propose various methods to configure the RIS phase shifts in outdoor communications, in order to optimize different communication utilities.
It is worth noting that accurate channel state information (CSI) is critical in optimizing the RIS parameters. Unfortunately, all the aforementioned work assumes perfect CSI without considering its acquisition difficulty. In fact, channel estimation in a RIS-assisted wireless system is much more challenging than in a conventional system. This is because the passive RIS elements are incapable of sensing and estimating channel information. Therefore, we shall rely on the receiver to estimate both the transmitter-to-RIS and RIS-to-receiver channels by observing only a noisy cascade of the two channels. 

To address the channel estimation challenge in RIS-assisted communication systems, some pioneering work has recently emerged.
 For example, Ref. \cite{LIS_ActiveCE} assumes that the RIS elements are fully active and are connected to a signal processing unit to perform channel estimation. Similarly, Ref. \cite{LIS_DL} requires a portion of the RIS elements to be active so that the channels of the passive elements can be inferred via a compressed-sensing based approach. Compared with active RIS elements, purely passive RIS elements are undoubtedly more appealing due to their extremely low hardware and deployment costs. Ref. \cite{LIS_CE1,LIS_CE2} show that channel estimation in passive-RIS-assisted systems can be converted into a sequence of conventional MIMO channel estimation problems by turning on one RIS element at a time. However, the training overhead of this method is proportional to the size of the RIS and may be prohibitively large as the RIS typically comprises a large number of elements. As such, the following question arises naturally: \emph{How to estimate the two cascaded channels with purely passive reflecting elements and an affordable training overhead?} Preliminary answers to this question appear in the recent papers \cite{LIS_He,JSACrevision_3,JSACrevision_1,JSACrevision_2}. In \cite{LIS_He}, the authors develop a cascaded channel estimation algorithm for a RIS-assisted \emph{single-user} MIMO system. Specifically, Ref. \cite{LIS_He} formulates the cascaded channel estimation problem as a combination of sparse matrix factorization and low-rank matrix completion by leveraging the programmable property of the RIS and the low-rankness of the propagation channel. {In \cite{JSACrevision_3}, the authors sequentially estimate the cascaded channels for users. Since users share the same RIS-to-receiver channel, the required training overhead is largely reduced by exploiting such channel correlations among users.}
{In \cite{JSACrevision_1}, the authors exploit the sparsity of the transmitter-to-RIS-to-receiver channels and estimate the cascaded channels based on compressed sensing.}
{In \cite{JSACrevision_2}, the authors employ parallel factor decomposition to alternatively estimate the cascaded channels.}

In this paper, we consider the cascaded channel estimation problem for an uplink RIS-assisted \emph{multiuser} MIMO system, where a fully passive RIS is used to assist the communication.
In practice, the RIS can be coated onto a wall, a ceiling, or a furniture in an indoor environment; or mounted on a building facade, an advertising panel, a traffic sign, or a highway fence in an outdoor environment. In a typical application scenario, both the base station (BS) and the RIS rarely move after deployment. As a result, the channel between the BS and the RIS can be modelled as a quasi-static end-to-end MIMO channel, in which most of the channel components evolve much more slowly compared with conventional mobile communication channels. Meanwhile, a small portion of the channel components of the BS-to-RIS channel may experience sudden changes. For example, an opening/closing of a door in an indoor scenario or a moving car in an outdoor scenario may change the scattering geometry. 
By modeling both the fast-varying and slow-varying channel components, we formulate the CSI acquisition problem as a \emph{matrix-calibration-based matrix factorization} task. Then, we propose a novel message-passing based algorithm to effectively estimate the two cascaded channels. Besides, we present an analytical framework to analyze the performance of the considered system. The main contributions of this paper are summarized as follows.
\begin{itemize}
	\item We characterize the slow-varying and fast-varying channel components of the RIS-to-BS channel by the Rician fading model. Unlike \cite{LIS_He} that assumes the low-rank channel matrix, we exploit the information on the slow-varying channel components for channel estimation. Specifically, we assume that the slow-varying channel components can be estimated by long-term channel averaging prior to the RIS channel estimation procedure. Based on this assumption, we formulate the cascaded channel estimation problem as a joint task of the RIS-to-BS channel matrix calibration and the RIS-to-user channel matrix estimation.
	\item We develop the posterior mean estimators under the Bayesian inference framework to infer the two cascaded channels and adopt the sum-product message passing algorithm to approximately compute the estimators. We further show that direct computation of the messages leads to prohibitively high complexity. To tackle this challenge, we introduce additional approximations to the messages based on the approximate message passing (AMP) framework \cite{MP_AMP1,MP_GAMP2,MP_BIGAMP1}. The proposed algorithm only needs to update the means and variances of the messages and hence avoids high-dimensional integrations in the canonical message passing algorithm. {Furthermore, different from \cite{JSACrevision_1} that exploits the sparsity of the joint transmitter-to-RIS-to-receiver channels, our work exploits the sparsity inherent in the separated transmitter-to-RIS and the RIS-to-receiver channels to enhance the estimation accuracy.}
	\item Based on the replica method from statistical physics \cite{SpinGlass}, we analyze the asymptotic performance of the posterior mean estimators in the large-system limit. We show that, with perfect knowledge of the channel prior distributions and under some appropriately chosen angle sampling bases, the mean square errors (MSEs) of the posterior mean estimators can be determined by the fixed point of a set of scalar equations. We further show by numerical simulations that the proposed approximate channel estimators can closely approach the replica method bound.
\end{itemize}

{As a final remark, we note that the approximations in our proposed algorithm share the same spirit with AMP. However, in this work, the cascaded channel coefficient matrices cannot be directly modelled with independent and identically distributed (\iid) elements, and shall be described by sparse representations under the domain transformed by angular sampling bases. In contrast, AMP \cite{MP_AMP1,MP_GAMP2} requires only one matrix to be unknown, and its extension to bilinear matrix factorization in \cite{MP_BIGAMP1} requires the two unknown matrices to be multiplied without domain transformations. Therefore, the existing work on AMP \cite{MP_AMP1,MP_GAMP2,MP_BIGAMP1} cannot be directly applied here.
}

The remainder of this paper is organized as follows. In Section \ref{sec_model}, we describe the RIS-assisted multiuser MIMO system model and the RIS channel estimation protocol. In Section \ref{sec3}, we discuss the channel model and formulate the cascaded channel estimation problem. In Section \ref{Sec_alg}, we develop the channel estimation algorithm and introduce additional approximations to reduce the computational complexity. In Section \ref{sec_ana}, we perform asymptotic analysis based on the replica method. Section \ref{Simulation} presents extensive numerical results of the proposed method, and the paper concludes in Section \ref{sec_conclusion}.

\emph{Notation}: Throughout, we use $\Real$ and $\Complex$ to denote the real and complex number sets, respectively. Regular letters, bold small letters, and bold capital letters are used to denote scalars, vectors, and matrices, respectively. We use $j\triangleq\sqrt{-1}$ to denote the imaginary unit. We use $(\cdot)^\star$, $(\cdot)^T$, and $(\cdot)^H$ to denote the conjugate, the transpose, and the conjugate transpose, respectively. We use $x_{ij}$ to denote the $(i,j)$-th entry of $\Xv$. We use $\Norm(\xv;\muv,\Sigmav)$ and $\CN(\xv;\muv,\Sigmav)$ to denote that $\xv$ follows the real normal and the circularly-symmetric normal distributions with mean $\muv$ and covariance $\Sigmav$, respectively. We use $\tr(\Xv)$ to denote the trace of $\Xv$, $\Iv$ to denote the identity matrix with an appropriate size, $\diag(\xv)$ to denote a diagonal matrix with the diagonal entries specified by $\xv$, and $\bf 1$ to denote the all-one vector with an appropriate size. We use $\norm{\cdot}_p$ to denote the $\ell_p$ norm, $\norm{\cdot}_F$ to denote the Frobenius norm, $\delta(\cdot)$ to denote the Dirac delta function, $\propto$ to denote equality up to a constant multiplicative factor, and $\E[\cdot]$ to denote the expectation operator.
\section{System Model}\label{sec_model}
\subsection{RIS-Assisted Multiuser MIMO} 
\begin{figure}[!t]
	\centering
	\includegraphics[width=3.3 in]{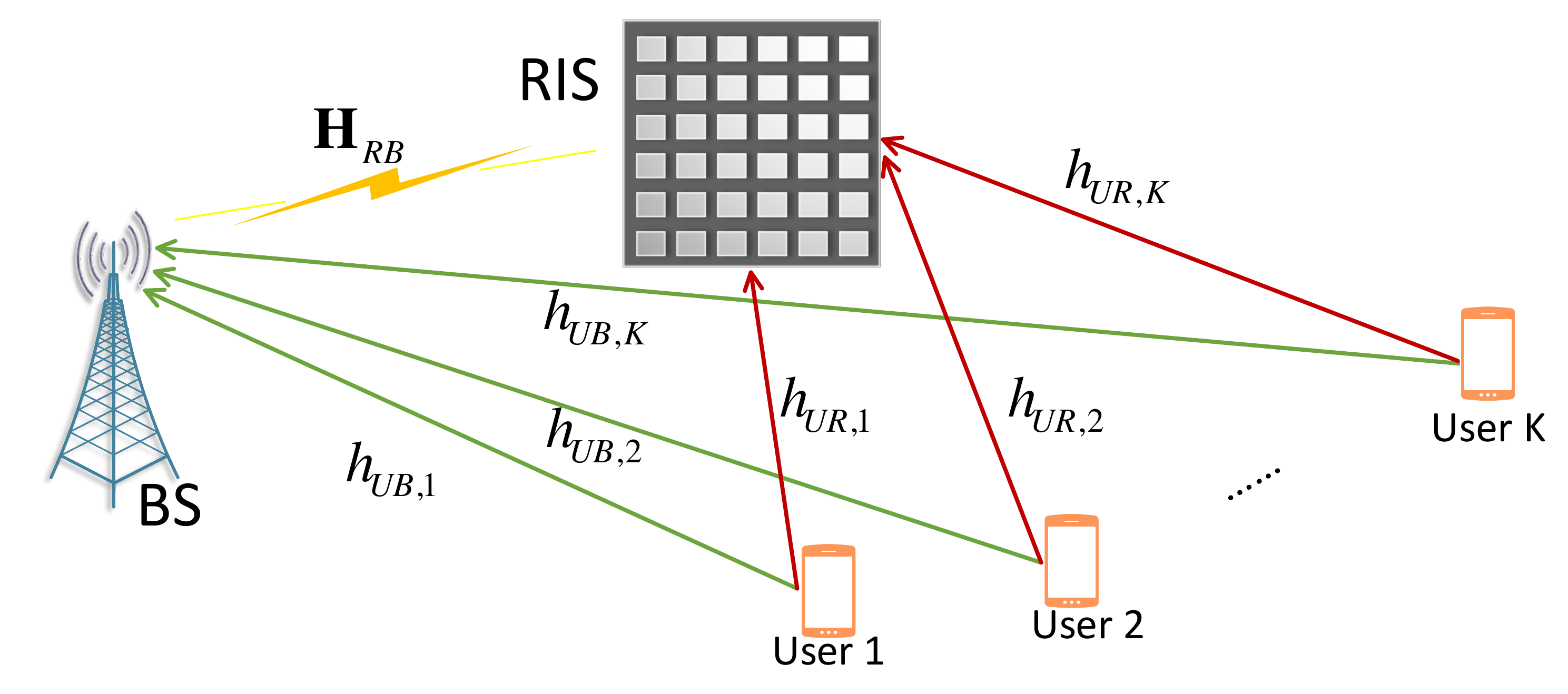}
	\caption{A RIS-assisted multiuser system.}
	\label{fig_system}
\end{figure}

Consider a single-cell RIS-assisted multiuser MIMO system depicted in Fig. \ref{fig_system}. Assume that $K$ single-antenna users simultaneously communicate with an $M$-antenna BS. A RIS comprising $L$ phase-shift elements is deployed to assist the communication between the users and the BS. A uniform linear array (ULA) is adopted at the BS, and the passive reflecting elements in the RIS are arranged in the form of an $L_1\times L_2$ uniform rectangular array (URA) with $L_1 L_2=L$. 

We assume a quasi-static flat fading channel model, where the channel coefficients remain invariant within the coherence time.
The coefficient vectors/matrix of the $k$-th-user-to-BS channel, the $k$-th-user-to-RIS channel, and the RIS-to-BS channel are denoted by $\hv_{UB,k}\in \Complex^{M\times 1}$, $\hv_{UR,k}\in \Complex^{L\times 1}$, and $\Hv_{RB}\in \Complex^{M\times L}$, respectively.
Moreover, we assume that the RIS elements induce independent phase shifts on the incident signals. We denote the RIS phase-shift vector at time $t$ by 
$\psiv(t)\triangleq[\varpi_{1}(t)e^{j\psi_{1}(t)},\varpi_{2}(t)e^{j\psi_{2}(t)},\cdots,\varpi_{L}(t)e^{j\psi_{L}(t)}]^T$,
where $\varpi_{l}(t) \in \{0,1\}$ represents the on/off state of the $l$-th RIS element, and $\psi_{l}(t)\in [0,2\pi)$ represents the phase shift of the $l$-th RIS element.  
\subsection{RIS Channel Estimation Protocol}
For the studied system, we aim to estimate the CSI of the channels  $\Hv_{RB}$, $\{\hv_{UR,k}\}$, and $\{\hv_{UB,k}\}$. We assume that all the direct channels $\{\hv_{UB,k}\}$ are accurately estimated at first and focus our discussions on estimating the RIS channels $\Hv_{RB}$ and $\{\hv_{UR,k}\}$.\footnote{By turning off the RIS reflecting elements, the estimation of $\{\hv_{UB,k}\}$ can be done by using conventional channel estimation methods for multiuser MIMO systems.}
To facilitate  RIS channel estimation, the users simultaneously transmit training sequences with length $T$ to the BS. Denote by $\xv_k=[x_{k1},\cdots,$ $x_{kT}]^T$ the training sequence of user $k$, where $x_{kt}$ is the training symbol of user $k$ in time slot $t$. We assume that the users transmit at constant power $\tau_X$, \ie $\E[\abs{x_{kt}}^2]=\tau_X, \forall k,t$.

Over the time duration $T$, all the RIS elements are turned on and are set to have the same phase shift. Without loss of generality, we assume that $\psiv(t)={\bf 1}, 1\leq t\leq T$.
Then, the received signal at the BS in time slot $t$ is given by
\begin{align}\label{eq1.1}
\yv_0(t)&=\sum_{k=1}^K (\hv_{UB,k}+\Hv_{RB}\hv_{UR,k})x_{kt}+\nv(t), 1\leq t \leq T,
\end{align}
where $\nv(t)$  is an additive white Gaussian noise (AWGN) vector following the distribution of $\CN(\nv(t);{\bf 0},\tau_N\Iv)$.

\section{RIS Channel Model and Problem Formulation}\label{sec3}
{In this section, we discuss the models of the RIS-to-BS channel matrix $\Hv_{RB}$ and the user-to-RIS channel vectors $\{\hv_{UR,k}\}$. Then, we show that the RIS channel estimation problem can be formulated as a sparse matrix factorization problem by employing virtual channel representations.}
\subsection{Channel Model}\label{sec_channelmodel}
	\begin{figure}[!t]
	\begin{minipage}[t]{\linewidth}
	\centering
	\includegraphics[width=2.8 in]{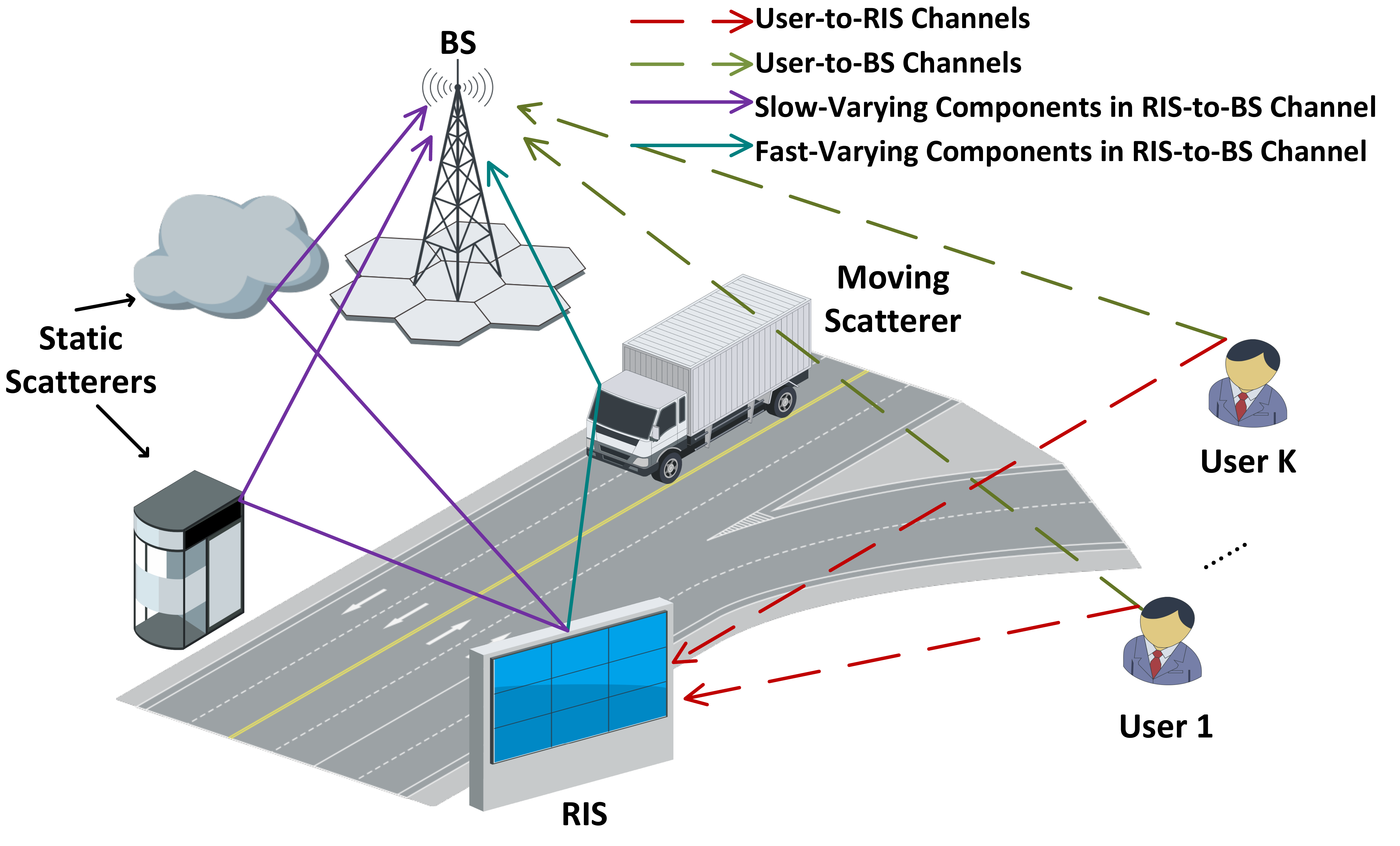}
	\caption{An example of the outdoor propagation geometry. The propagation path through a moving truck belongs to the fast-varying channel components between the BS and the RIS, while the paths from static scatterers belong to the slow-varying channel components.}
	\label{fig_fig2}
	\end{minipage}
	\begin{minipage}[t]{\linewidth}
		\centering
	\includegraphics[width=3.3 in]{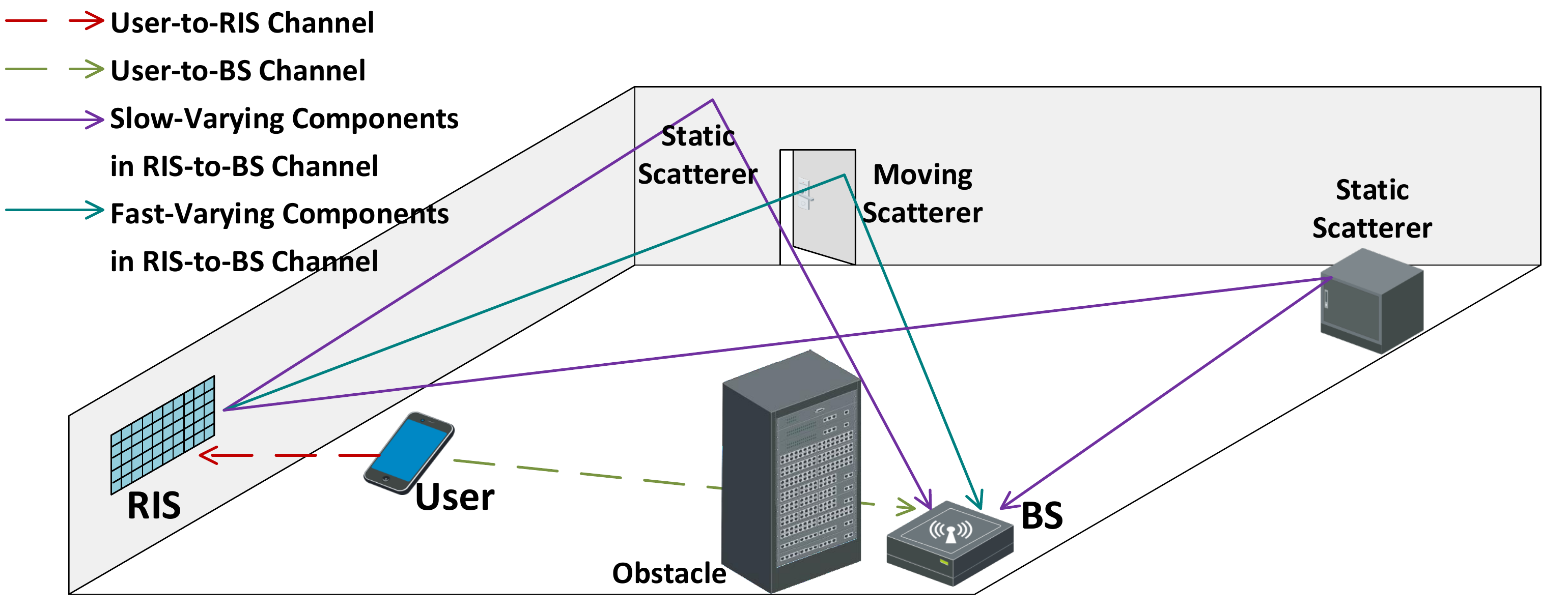}
	\caption{An example of the indoor propagation geometry. The propagation path through a closing door belongs to the fast-varying channel components between the BS and the RIS, while the paths from static scatterers belong to the slow-varying channel components.}
	\label{fig_fig3}
	\end{minipage}
\end{figure}
{Consider the RIS-to-BS channel $\Hv_{RB}$.
On one hand, since the BS and the RIS rarely move after deployment, a majority of the channel paths evolve very slowly, as compared with the channel coherence time (determined by the movement of mobile ends). We refer to these paths as the slow-varying channel components of $\Hv_{RB}$, denoted by $\Hvb_{RB}$. On the other hand, a small portion of the propagation paths may experience fast changes due to the change of the propagation geometry. We refer to them as the fast-varying channel components of $\Hv_{RB}$, denoted by $\Hvt_{RB}$. To illustrate this, consider an outdoor scenario depicted in Fig. \ref{fig_fig2}, where a RIS is mounted on an advertising panel by the road. The propagation paths from static scattering clusters between the BS and the RIS changes very slowly and $\Hvb_{RB}$ sums up the paths from all the static scattering clusters. Meanwhile, suppose, for example, that a truck is moving past the RIS. The path through the truck evolves fast, as the corresponding propagation geometry is quickly changing. This fast-varying path is modelled in $\Hvt_{RB}$. As another example, consider an indoor environment depicted in Fig. \ref{fig_fig3}, where a RIS is coated on a wall to assist the communication between the user and the BS. Most of the propagation paths are from static scattering clusters and are modelled in the slow-varying component matrix $\Hvb_{RB}$. Meanwhile, the paths through non-static scatterers (\eg the closing door in Fig. \ref{fig_fig3}) are modelled in the fast-varying component matrix $\Hvt_{RB}$.}

To capture $\Hvb_{RB}$ and $\Hvt_{RB}$, we model $\Hv_{RB}$ by the MIMO Rician fading model as \cite{MIMO_DTse}
\begin{align}\label{Hrmodel0}
\Hv_{RB}=\sqrt{\frac{\kappa}{\kappa+1}}  \Hvb_{RB}+\sqrt{\frac{1}{\kappa+1}}\Hvt_{RB},
\end{align}
where 
$\kappa$ is the Rician factor denoting the power ratio between $\Hvb_{RB}$ and $\Hvt_{RB}$.

By specifying the paths in the prorogation environment, $\Hvb_{RB}$ and $\Hvt_{RB}$ are given by
\begin{align}\label{Hrbmodel0}
&\Hvb_{RB}=\sqrt{\beta_0}\sum_{p=1}^{\bar P_{RB}}  \a_{p} \av_{B}(\theta_{p})\av_{R}^H(\phi_{p},\sigma_{p}),\\
 &\Hvt_{RB}=\sqrt{\beta_0}\sum_{p=1}^{\widetilde P_{RB}}  \a_{p} \av_{B}(\theta_{p})\av_{R}^H(\phi_{p},\sigma_{p}),\label{Hrtmodel0}
\end{align}
where $\beta_0$ is the large-scale path gain encompassing distance-dependent path loss and shadowing; $\bar P_{RB}$ (or $\widetilde P_{RB}$) is the number of the slow-varying (or fast-varying) paths; $\a_{p}$ is the corresponding complex-valued channel coefficient of the $p$-th path; $\theta_{p}$ is the corresponding azimuth angle-of-arrival (AoA) to the BS; $\phi_{p}$ (or $\sigma_{p}$) is  the corresponding azimuth (or elevation) angle-of-departure (AoD) from the RIS;
and $\av_{B}$ (or $\av_{R}$) is the steering vector associated with the BS (or RIS) antenna geometry. Specifically, they are given by
\begin{subequations}
\begin{align}
\label{ULA}
&\av_{B}(\theta)=\fv_M(\sin(\theta)),\\
\label{URA}
&\av_{R}(\phi,\sigma)=\fv_{L_2}(-\cos(\sigma)\cos(\phi))\!\otimes
\!\fv_{L_1}(\cos(\sigma)\sin(\phi)),
\end{align}
\end{subequations}
where $\otimes$ denotes the Kronecker product; and
\begin{align}\label{temp310}
	\fv_N(x)\triangleq\frac{1}{\sqrt{N}}\left[1,e^{-j\frac{2\pi}{\varrho}dx},\cdots,e^{-j\frac{2\pi}{\varrho}d(N-1)x}\right]^T.
\end{align}
In \eqref{temp310}, $\varrho$ denotes the carrier wavelength; and $d$ denotes the distance between any two adjacent antennas. Here, we set $d/\varrho=1/2$ for simplicity.

Similarly to \eqref{Hrbmodel0}--\eqref{Hrtmodel0}, we represent $\hv_{UR,k}$ as
 \begin{align}
&\hv_{UR,k}=\sqrt{\beta_k}\sum_{p=1}^{ P_{k}}  \a_{p} \av_{R}(\phi_{p},\sigma_{p}), \label{Humodel0} 
\end{align}
where 
$\beta_k$ is the large-scale path gain for the channel between the $k$-th user and the RIS;
and $P_k$ is the number of paths between the $k$-th user and the RIS. 


Finally, we assume without loss of generality that 
{$\E[\norm{\Hvb_{RB}}^2_F]=\E[\norm{\Hvt_{RB}}^2_F]=\beta_0 ML$ and $\E[\norm{\hv_{UR,k}}^2_2]=\beta_k L, \forall k$.}
In other words, $\beta_0$ (or $\beta_k$) can be regarded as the average power attenuation for each RIS-BS (or user-RIS) antenna pair.
\subsection{Virtual Channel Representation}
We assume that the slow-varying component matrix $\Hvb_{RB}$ in \eqref{Hrbmodel0} keeps static over a time interval much larger than the coherence block length. As a consequence, $\sqrt{\kappa/(\kappa+1)}\Hvb_{RB}$ can be accurately estimated by long-term channel averaging prior to the RIS channel estimation procedure.\footnote{The assumption of the perfect knowledge of $\sqrt{\kappa/(\kappa+1)}\Hvb_{RB}$ does not lose any generality since any possible error in the channel averaging process can be absorbed into $\sqrt{1/(\kappa+1)}\Hvt_{RB}$.}

{Furthermore, as illustrated above, the fast-varying component matrix $\Hvt_{RB}$ contains a limited number of paths, \ie $\widetilde P_{RB}$ in \eqref{Hrtmodel0} is small.} Following  \cite{MIMO_JFang}, we employ a pre-discretized sampling grid $\varthetav$ with length $M^\prime$ ($\geq M$) to discretize $\{\sin(\theta_p)\}_{1\leq p \leq \widetilde P_{RB}}$ over $[0,1]$.
Similarly, we employ two sampling grids $\varphiv$ with length $L_1^\prime$ ($\geq L_1$) and $\varsigmav$ with length $L_2^\prime$ ($\geq L_2$) to discretize $\{\cos\left( \sigma_p\right) \sin\left( \phi_p\right) \}_{1\leq p \leq \widetilde P_{RB}}$ and $\{-\cos\left( \sigma_p\right) \cos\left( \phi_p\right)\}_{1\leq p \leq \widetilde P_{RB}} $, respectively.
Then, we represent the fast-varying component matrix $\Hvt_{RB}$ under the angular bases as \cite{MIMO_JFang}
\begin{align}\label{Hrmodel}
\sqrt{\frac{1}{\kappa+1}}\Hvt_{RB}=\Av_{B}\Sv\left(\underbrace{\Av_{R,v}\otimes \Av_{R,h}}_{=\Av_{R}}\right)^H,
\end{align}
where $\Av_{B}\triangleq [\fv_M(\vartheta_{1}),\cdots,\fv_M(\vartheta_{M^\prime})]$ is an over-complete array response with $\fv_M$ defined in \eqref{temp310};
$\Av_{R,h}\triangleq [\fv_{L_1}(\varphi_{1}),\cdots,\fv_{L_1}(\varphi_{L_1^\prime})]$ (or $\Av_{R,v}\triangleq [\fv_{L_2}(\varsigma_{1}),$ $\cdots,\fv_{L_2}(\varsigma_{L_2^\prime})]$) is an over-complete horizontal (or vertical) array response; and $\Sv\in  \Complex^{M^\prime \times L^\prime}$ is the corresponding channel coefficient matrix in the angular domain with $L^\prime=L_1^\prime  L_2^\prime$. {Note that the $(i,j)$-th entry of $\Sv$ corresponds to the channel coefficient of $\Hvt_{RB}$ along the path specified by the $i$-th AoA steering vector in $\Av_B$ and the $j$-th AoD steering vector in $\Av_R$. Since the total number of paths $\widetilde P_{RB}$ is small, only a few entries of $\Sv$ are nonzero with each corresponding to a channel path. That is, $\Sv$ is a sparse matrix.}

Similarly to \eqref{Hrmodel}, $\hv_{UR,k}$ can be represented as
\begin{align}\label{Humodel}
\hv_{UR,k}=\Av_{R}\gv_{k},
\end{align}
where  $\gv_{k}\in  \Complex^{L^\prime\times 1}$ represents the channel coefficients of $\hv_{UR,k}$ in the angular domain.
Experimental studies have shown that the propagation channel often exhibits limited scattering geometry \cite{Channel_Geometry}. As a consequence, $\gv_k$ is also sparse, where each nonzero value corresponds to a channel path.
 We note that the sparsity of $\Sv$ and $\{\gv_k\}$ plays an important role in our channel estimation design.
\subsection{Cascaded Channel Estimation}
With the above channel representations, we can rewrite \eqref{eq1.1} in a matrix form, as shown in the following proposition.
\proposition
{With \eqref{Hrmodel0}, \eqref{Hrmodel}, and \eqref{Humodel}, the model in \eqref{eq1.1} is equivalent to 
\begin{align}\label{eq2}
\Yv
&=\left( \Hv_0+\Av_{B}\Sv\Rv\right) \Gv\Xv+\Nv,
\end{align}
where $\Xv=[\xv_1,\cdots,\xv_K]^T$; $\Nv=[\nv(1),\cdots,\nv(T)]$, $\Yv=[\yv_0(1),\cdots,\yv_0(T)]-[\hv_{UB,1},\cdots,\hv_{UB,K}]\Xv$; $\Hv_0\triangleq \sqrt{{\kappa}/( \kappa+1)} \Hvb_{RB}\Av_R\in  \Complex^{M\times L^\prime}$; $\Rv\triangleq \Av_R^H\Av_R\in  \Complex^{ L^\prime\times L^\prime}$; and $\Gv\triangleq[\gv_{1},\cdots,\gv_{K}]\in  \Complex^{L^\prime\times K}$.
\label{pro1}}
\IEEEproof{See Appendix \ref{appa0}.}

Upon the reception of $\Yv$ in \eqref{eq2}, the BS aims to factorize the channel matrices $\Sv$ and $\Gv$ with the knowledge of the training signal matrix $\Xv$. Once the angular bases $\Av_B$ and $\Av_R$ are predetermined and the slow-varying component matrix $\sqrt{\kappa/(\kappa+1)}\Hvb_{RB}$ is given, the sensing matrices $\Av_B$, $\Hv_0$, and $\Rv$ are also known to the BS. We refer to the above problem as \emph{matrix-calibration} based cascaded channel estimation, since this problem bears some similarity to the {blind matrix calibration (BMC)} problem \cite{Phase_ISIT}.
{\remark{The BMC problem in \cite{Phase_ISIT} is defined as to calibrate $\Av$ and estimate $\Bv$ from a noisy observation of the product $\Av\Bv$, where the dictionary $\Av$ is partially known. That is, $\Av=\Av_0+\Av_1$ with the known part $\Av_0$ and the unknown perturbation $\Av_1$. 
Rigorously speaking, the problem in  \eqref{eq2} is not exactly a BMC problem. The difference is two-fold: {First}, in  \eqref{eq2}, the observation of the matrix product $\Zv=\left( \Hv_0+\Av_{B}\Sv\Rv\right) \Gv$ is through a linear system $\Yv=\Zv\Xv+\Nv$. In BMC, $\Xv$ is limited to $\Xv=\Iv$. {Second}, BMC often assumes an \iid Gaussian distribution for perturbation $\Av_1$ \cite{Phase_ISIT,MP_BIGAMP3}, whereas we exploit the hidden sparsity of the perturbation matrix under the domain transformed by $\Av_B$ and $\Rv$ in \eqref{eq2}. As aforementioned, such a sparse representation of the perturbation models the fast channel variation
in the RIS-to-BS channel.
}}

\section{Matrix-Calibration Based Cascaded Channel Estimation Algorithm}\label{Sec_alg}
In this section, we first derive the posterior mean estimators of $\Sv$ and $\Gv$ given $\Yv$ in \eqref{eq2} under the Bayesian inference framework.
We then resort to sum-product message passing to compute the estimators. To reduce the computational complexity, we impose additional approximations to simplify the message updates in the large-system limit. Finally, we analyze the computational complexity of the proposed algorithm. 

\subsection{Bayesian Inference}\label{sec_bi}
Define $ \Wv\triangleq\Hv_0+\Av_{B}\Sv\Rv$, $\Zv\triangleq\Wv\Gv$, and $\Qv\triangleq\Zv\Xv$. Under the assumption of AWGN, we have 
\begin{equation}
\label{gout}
p(\Yv|\Qv)=\prod_{m=1}^M\prod_{t=1}^T  \CN\left( y_{mt};q_{mt},\tau_N\right) .
\end{equation}
Motivated by the sparsity of $\Sv$ and $\Gv$, we employ Bernoulli-Gaussian distributions to model their prior distributions as
\begin{align}
\label{gin1}
p(\Sv)&=\prod_{m^\prime=1}^{M^\prime}\prod_{l^\prime=1}^{L^\prime}
(1-\lambda_S)\delta(s_{m^\prime l^\prime})+\lambda_S\CN\left(s_{m^\prime l^\prime};0,\tau_S \right),\\
\label{gin2}
p(\Gv)& =\prod_{l=1}^{L^\prime}\prod_{k=1}^{K} (1-\lambda_G)\delta(g_{lk})+\lambda_G\CN\left(g_{lk};0,\tau_{G} \right),
\end{align}
where $\lambda_S$ (or $\lambda_G$) is the corresponding Bernoulli parameter of $\Sv$ (or $\Gv$); and $\tau_S$ (or $\tau_{G}$) is the variance of the nonzero entries of $\Sv$ (or $\Gv$).

{With the prior distributions \eqref{gout}--\eqref{gin2}, the following proposition states the minimum MSEs (MMSEs) of $\Sv$ and $\Gv$.}
\proposition{The posterior distribution is given by
\begin{align}\label{post}
p(\Sv,\Gv|\Yv)=\frac{1}{p(\Yv)} p(\Yv|\Sv,\Gv)p(\Sv)p(\Gv),
\end{align}
where $p(\Yv)=\int p(\Yv|\Sv,\Gv)p(\Sv)p(\Gv)\mathrm{d}\Sv\mathrm{d}\Gv$. 
\begin{table}[!t]
	\centering
	\caption{Notation of factor nodes}\label{BPnotations}
	\begin{tabular}{l|l|l}
		\toprule
		Factor& Distribution & Exact Form\\
		\midrule
				$p(s_{m^\prime l^\prime})$&$p(s_{m^\prime l^\prime})$&$(1-\lambda_S)\delta(s_{m^\prime l^\prime})+$\\
				&&$\lambda_S\CN\left(s_{m^\prime l^\prime};0,\tau_S\right)$\\
		$p(g_{lk})$&$p(g_{lk})$&$(1-\lambda_G)\delta(g_{lk})+$\\
		&&$\lambda_G\CN\left(g_{lk};0,\tau_{G} \right)$\\
			$ws_{ml}$&$p(w_{ml}|s_{m^\prime l^\prime}, \forall m^\prime, l^\prime)$&$\delta(w_{ml}\!-\!h_{0,ml}-$\\
			&&$\!\sum_{m^\prime,l^\prime}a_{B,mm^\prime}s_{m^\prime l^\prime}r_{l^\prime l})$\\
		$zwg_{mk}$&$p(z_{mk}|w_{ml},g_{lk}, \forall l)$&$\delta(z_{mk}-\sum_{l=1}^{L^\prime}w_{ml}g_{lk})$\\
			$qz_{mt}$&$p(q_{mt}|z_{mk}, \forall k)$&$\delta(q_{mt}- \sum_{k=1}^{K}z_{mk}x_{kt})$\\
		$p(y_{mt}|q_{mt})$&$p(y_{mt}|q_{mt})$&$\CN(y_{mt};q_{mt},\tau_N)$\\
		\bottomrule
	\end{tabular}
\end{table}
Furthermore, the MMSEs of $\Sv$ and $\Gv$ are given by
\begin{subequations}\label{mseMatrix}
\begin{align}
&\text{MMSE}_\Sv=\frac{1}{M^\prime L^\prime}\E\left[\norm{\Sv-\Svh}^2_F\right],\label{mseS}\\
&\text{MMSE}_\Gv=\frac{1} {L^\prime K}\E\left[\norm{\Gv-\Gvh}^2_F\right],\label{mseG}
\end{align}
\end{subequations}
where the expectations are taken over the joint distribution of $\Sv$, $\Gv$, and $\Yv$; and $\Svh=[\sh_{m^\prime l^\prime}]$ (or $\Gvh=[\gh_{l k}]$) is the posterior mean estimator of $\Sv$ (or $\Gv$), given by
\begin{align}
	\sh_{m^\prime l^\prime}&=\int s_{m^\prime l^\prime} p(s_{m^\prime l^\prime}|\Yv) \mathrm{d} s_{m^\prime l^\prime}, \gh_{l k}=\int g_{lk} p(g_{lk}|\Yv) \mathrm{d} g_{lk}.\label{temp1000}
	\end{align}
In the above, $p(s_{m^\prime l^\prime}|\Yv)=\int\int p(\Sv,\Gv|\Yv) \mathrm{d}{\Gv}\mathrm{d}(\Sv\setminus s_{m^\prime l^\prime})$ and $ p(g_{lk}|\Yv)=\int\int p(\Sv,\Gv|\Yv)\mathrm{d}\Sv\mathrm{d}(\Gv\setminus g_{lk})$ are the marginal distributions with respect to $s_{m^\prime l^\prime}$ and $g_{lk}$, respectively, where $\Xv\setminus x_{ij}$ means the collection of the elements of matrix $\Xv$ except for the $(i,j)$-th one.}
\IEEEproof{The result in \eqref{post} follows from Bayes' theorem, and \eqref{temp1000} follows from \cite[Page 143]{HVPoor}.
}

	
Exact evaluation of $\Svh$ and $\Gvh$ are generally intractable due to the high-dimensional integrations involved in the marginalization. In the following, we provide an approximate solution by following the message passing principle.
\subsection{Message Passing for Marginal Posterior Computation}
Plugging \eqref{gout}--\eqref{gin2} into \eqref{post}, we obtain
\begin{align}\label{post2}
  &p(\Sv,\Gv|\Yv)=\frac{1}{p(\Yv)}\!
  \left(\prod_{m=1}^M\prod_{t=1}^T p(y_{mt}|q_{mt})p(q_{mt}|z_{mk}, \forall k) \right) \nonumber\\
  &\left( \prod_{m=1}^M\prod_{k=1}^Kp(z_{mk}|w_{ml},g_{lk}, 1\leq l \leq L^\prime) \right)\!\left(\prod_{l=1}^{L^\prime}\prod_{k=1}^{K} p(g_{lk}) \right)\nonumber\\
  & \left(\prod_{m=1}^{M}\prod_{l=1}^{L^\prime}p(w_{ml}|s_{m^\prime l^\prime}, \forall m^\prime, l^\prime)\right)\!\left(\prod_{m^\prime=1}^{M^\prime}\prod_{l^\prime=1}^{L^\prime} p(s_{m^\prime l^\prime}) \right)\!,
\end{align}
where the factorizable distributions are defined in Table \ref{BPnotations}. 
\begin{figure}[!t]
	\centering
	\includegraphics[width=3.5 in]{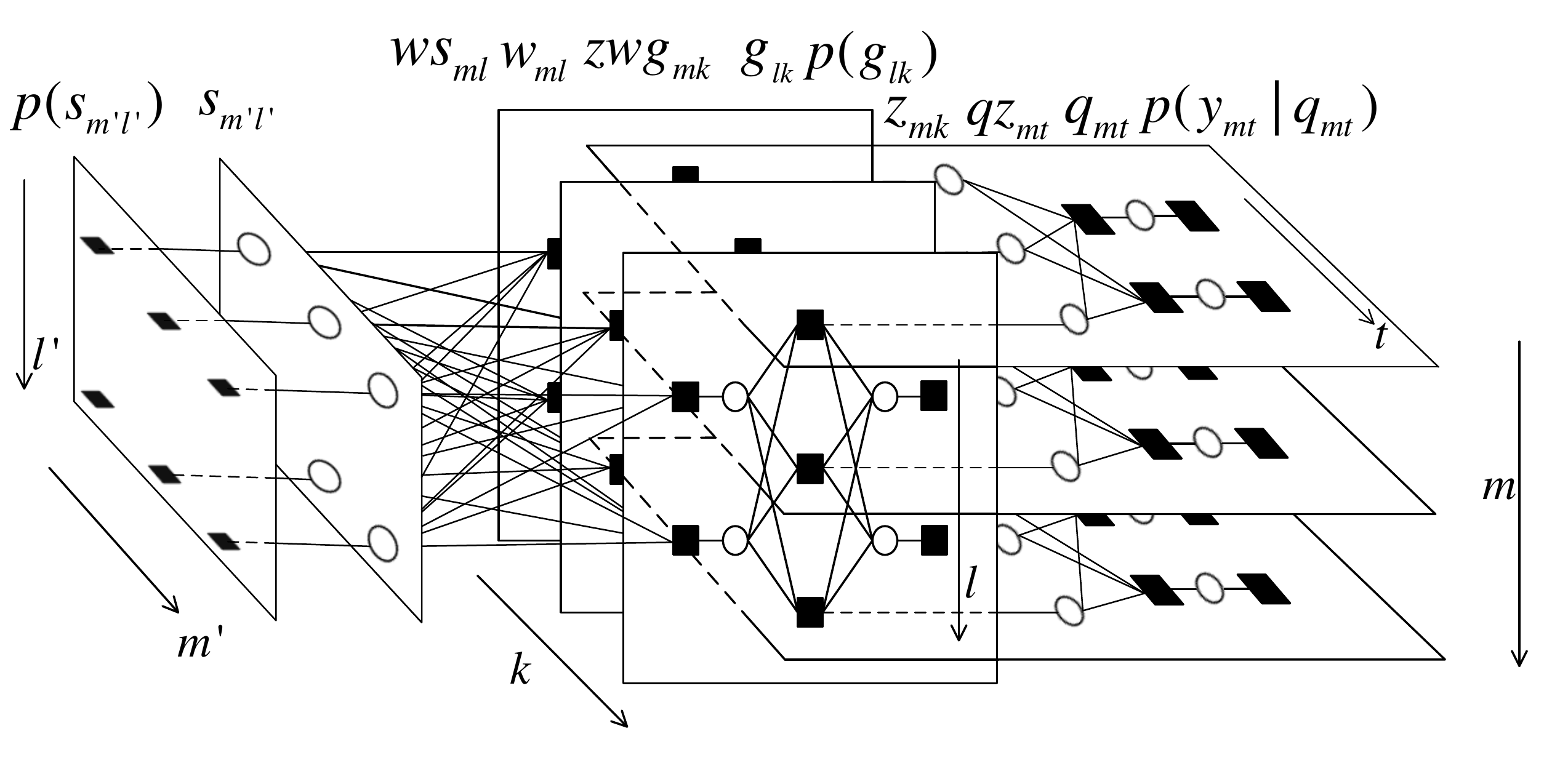}
 	\caption{An illustration of the factor graph representation for $M=M^\prime=K=3$ and $T=L^\prime=2$, where blank circles and black squares represent variable nodes and factor nodes, respectively.}
	\label{fig_fg}
\end{figure}

We construct a factor graph to represent \eqref{post2} and apply the canonical message passing algorithm to approximately compute the estimators in \eqref{temp1000}. The factor graph is depicted in Fig. \ref{fig_fg}. The variables $\Sv$, $\Gv$, $\Wv$, $\Zv$, and $\Qv$ are represented by the variable nodes $\{s_{m^\prime l^\prime}\}_{1\leq m^\prime \leq M^\prime, 1\leq l^\prime\leq L^\prime}$, $\{g_{lk}\}_{1\leq l \leq L^\prime, 1\leq k\leq K}$, $\{w_{ml}\}_{1\leq m \leq M, 1\leq l\leq L^\prime}$, $\{z_{mk}\}_{1\leq m \leq M, 1\leq k\leq K}$, and
$\{q_{mt}\}_{1\leq m \leq M, 1\leq t\leq T}$, respectively.
The factorizable pdfs in \eqref{post2}, represented by factor nodes $\{p(s_{m^\prime l^\prime})\}$, $\{p(g_{lk})\}$, $\{ws_{ml}\}$, $\{zwg_{mk}\}$, $\{qz_{mt}\}$, and $\{p(y_{mt}|q_{mt})\}$,
are connected to their associated arguments. We summarize the notation of the factor nodes in Table \ref{BPnotations}. 
Denote by $\Delta^i_{a \to b}(\cdot)$ the message from node $a$ to $b$ in iteration $i$, and by $\Delta_{c}^i(\cdot)$ the marginal message computed at variable node $c$ in iteration $i$. Applying the sum-product rule, we obtain the following messages:
\subsubsection{Messages between $\{qz_{mt}\}$ and $\{z_{mk}\}$} For $1\leq m \leq M, 1\leq t\leq T, 1\leq k\leq K$,
\begin{align}
\Delta_{qz_{mt}\to z_{mk}}^i (z_{mk})&\propto \int \CN\left(y_{mt};\sum_{k=1}^{K} z_{mk}x_{kt},\tau_N \right)\nonumber\\
&\times\prod_{j  \neq k} \left( \Delta^i_{z_{mj}\to qz_{mt}}(z_{mj})\mathrm{d} z_{mj}\right) ,   \label{qtog}\\
\Delta_{ z_{mk}\to qz_{mt}}^{i+1}(z_{mk})&\propto \mathcal{P}^i_{z_{mk}}(z_{mk}) \prod_{j \neq t}\Delta^i_{qz_{mj}\to z_{mk}}(z_{mk})\label{gtoq},
\end{align}
where the auxiliary distribution $\mathcal{P}^i_{z_{mk}}(z_{mk})$ is defined as
\begin{align}
&\mathcal{P}^i_{z_{mk}}(z_{mk})\propto\int p(z_{mk}|w_{ml},g_{lk}, 1\leq l \leq L^\prime) 
\nonumber\\
&\times\prod_{l=1}^{L^\prime}\left(  \Delta^i_{w_{ml}\to zwg_{mk}}(w_{ml})\Delta^i_{g_{lk}\to zwg_{mk}}(g_{lk})\mathrm{d}w_{ml}\mathrm{d}g_{lk} \right).\label{pz}
\end{align}
 
\subsubsection{Messages between $\{g_{lk}\}$ and $\{zwg_{mk}\}$} For $1\leq l \leq L^\prime, 1\leq m \leq M, 1\leq k\leq
K$,
\begin{align}
&\Delta_{zwg_{mk}\to g_{lk}}^i(g_{lk})\propto  \int \mathrm{d} z_{mk}p(z_{mk}|w_{ml},g_{lk}, 1\leq l \leq L^\prime)
 \nonumber\\
& \times\prod_{t=1}^T\Delta_{qz_{mt}\to z_{mk}}^i (z_{mk})\prod_{j \neq l} \left( \Delta_{g_{jk}\to zwg_{mk}}^i(g_{jk})\mathrm{d} g_{jk}\right) \nonumber\\
& \times\prod_{l=1}^{L^\prime}\left( \Delta_{w_{ml}\to zwg_{mk}}^i(w_{ml})\mathrm{d} w_{ml}\right),  \label{gtou}\\
&\Delta_{g_{lk} \to  zwg_{mk}}^{i+1}(g_{lk})\propto p(g_{lk}) \prod_{j \neq  m}\Delta_{zwg_{jk}\to g_{lk}}^i(g_{lk}).\label{utog}
\end{align}
\subsubsection{Messages between $\{w_{ml}\}$ and $\{zwg_{mk}\}$} For $1\leq l \leq L^\prime, 1\leq m \leq M, 1\leq k\leq K$,
\begin{align}
&\Delta_{zwg_{mk}\to w_{ml}}^i(w_{ml})\propto  \int \mathrm{d}z_{mk}p(z_{mk}|w_{ml},g_{lk}, 1\leq l \leq L^\prime) 
\nonumber\\
&\times \prod_{t=1}^T\Delta_{qz_{mt}\to z_{mk}}^i (z_{mk})\prod_{l=1}^{L^\prime} \left( \Delta_{g_{lk}\to zwg_{mk}}^i(g_{lk})\mathrm{d}g_{lk}\right)  \nonumber\\
&\times\prod_{j \neq l}\left( \Delta_{w_{mj}\to zwg_{mk}}^i(w_{mj})\mathrm{d}w_{mj}\right), \label{gtow}\\
&\Delta_{w_{ml} \to  zwg_{mk}}^{i+1}(w_{ml})\propto \mathcal{P}^i_{w_{ml}}(w_{ml}) \prod_{j \neq  k}\Delta_{ zwg_{mj}\to w_{ml}}^i(w_{ml}),\label{wtog}
\end{align}
where 
\begin{align}
\mathcal{P}^i_{w_{ml}}(w_{ml})\propto\int& \prod_{m^\prime=1}^{M^\prime}\prod_{l^\prime=1}^{L^\prime}\left( \Delta^i_{s_{m^\prime l^\prime}\to ws_{ml}}(s_{m^\prime l^\prime})\mathrm{d}s_{m^\prime l^\prime}\right) \nonumber\\
&\times p(w_{ml}|s_{m^\prime l^\prime}, \forall m^\prime, l^\prime).\label{pw}
\end{align}
\subsubsection{Messages between $\{s_{m^\prime l^\prime}\}$ and $\{ws_{ml}\}$}For $1\leq l,l^\prime \leq L^\prime, 1\leq m^\prime \leq M^\prime, 1\leq m \leq M, $
\begin{align}
&\Delta_{ws_{ml}\to s_{m^\prime l^\prime}}^i(s_{m^\prime l^\prime})\propto \int\mathrm{d}w_{ml}p(w_{ml}|s_{m^\prime l^\prime}, \forall m^\prime, l^\prime)\nonumber\\
& \times\prod_{k=1}^{K} \Delta_{zwg_{mk}\to w_{ml}}^i(w_{ml})\!\prod_{(j,n)\neq (m^\prime\!,l^\prime)}\!\left(\Delta_{s_{jn} \to  ws_{ml}}^{i}(s_{jn}) \mathrm{d}s_{jn}\right) \label{ftor}\!,\\
&\Delta_{s_{m^\prime l^\prime} \to  ws_{ml}}^{i+1}(s_{m^\prime l^\prime})\propto p(s_{m^\prime l^\prime})\prod_{(j,n)\neq (m,l)} \Delta_{ws_{jn}\to s_{m^\prime l^\prime}}^i(s_{m^\prime l^\prime}).\label{rtof}
\end{align}
\subsubsection{Marginal messages at variable nodes} For $1\leq l,l^\prime \leq L^\prime, 1\leq m^\prime \leq M^\prime, 1\leq m \leq M, 1\leq k\leq K$,  
\begin{align}
&\Delta_{ z_{mk}}^{i+1}(z_{mk})\propto \mathcal{P}^i_{z_{mk}}(z_{mk}) \prod_{t=1}^T\Delta^i_{qz_{mt}\to z_{mk}}(z_{mk})\label{zpost},\\
&\Delta_{w_{ml}}^{i+1}(w_{ml})\propto \mathcal{P}^i_{w_{ml}}(w_{ml}) \prod_{k=1}^{K}\Delta_{ zwg_{mk}\to w_{ml}}^i(w_{ml}),\label{wpost}\\
&\Delta_{g_{lk}}^{i+1}(g_{lk})\propto p(g_{lk}) \prod_{m=1}^M\Delta_{ zwg_{mk}\to g_{lk}}^i(g_{lk}),\label{upost}\\
&\Delta_{s_{m^\prime l^\prime}}^{i+1}(s_{m^\prime l^\prime})\propto p(s_{m^\prime l^\prime})\prod_{m=1}^{M}\prod_{l=1}^{L^\prime} \Delta_{ws_{ml}\to s_{m^\prime l^\prime}}^i(s_{m^\prime l^\prime}).\label{rpost}
\end{align}
\subsection{Approximations for Message Passing}\label{sec_33}
\begin{table}[!t]
	\centering
	\caption{Notations of means and variances for messages}\label{table1.5}
	\begin{tabular}{l|l|l}
		\toprule
		Message& Mean & Variance\\
		\midrule
		$\Delta_{g_{lk}\to  zwg_{mk}}^{i}(g_{lk})$&$\gh_{lk,m}(i)$&$v^g_{lk,m}(i)$\\
		$\Delta_{s_{m^\prime l^\prime} \to  ws_{ml}}^{i}(s_{m^\prime l^\prime})$&$\sh_{m^\prime l^\prime,ml}(i)$&$v^s_{m^\prime l^\prime,ml}(i)$\\
		$ \Delta^i_{w_{ml}\to zwg_{mk}}(w_{ml})$&$\wh_{ml,k}(i)$&$v^w_{ml,k}(i)$\\
		$\Delta_{ z_{mk}\to qz_{mt}}^{i}(z_{mk})$&$\zh_{mk,t}(i)$&$v^z_{mk,t}(i)$\\
		$\Delta_{g_{lk}}^{i}(g_{lk})$&$\gh_{lk}(i)$&$v^g_{lk}(i)$\\
		$\Delta_{s_{m^\prime l^\prime}}^{i}(s_{m^\prime l^\prime})$&$\sh_{m^\prime l^\prime}(i)$&$v^s_{m^\prime l^\prime}(i)$\\
				$\Delta_{w_{ml}}^{i}(w_{ml})$&$\wh_{ml}(i)$&$v^w_{ml}(i)$\\
			$\Delta_{z_{mk}}^i (z_{mk})$&$\zh_{mk}(i)$&$v^z_{mk}(i)$\\
		\bottomrule
	\end{tabular}
\end{table}
The messages in \eqref{qtog}--\eqref{rpost} are computationally intractable in general due to the high-dimensional integrations and normalizations therein. To tackle this, we simplify the calculation of \eqref{qtog}--\eqref{rpost} by following the idea of AMP \cite{MP_AMP1,MP_GAMP2} in the large-system limit, \ie $M,M^\prime,K,L,L^\prime,T,\tau_N\to \infty$ with the ratios $M/K$, $M^\prime/K$, $L/K$, $L^\prime/K$, $T/K$, and $\tau_N/K^2$ fixed.
For ease of notation, we define the means and variances of the messages as in Table \ref{table1.5}.
In the following, we sketch the main approximations involved in the algorithm design, while the detailed derivations can be found in Appendix \ref{appa}. 
\begin{itemize}
	\item A second-order Taylor expansion is adopted to approximate $\mathcal{P}^i_{z_{mk}}$ as a Gaussian distribution. Moreover, the CLT argument \cite{MP_AMP1} approximates $\prod_{j  \neq k} \Delta^i_{z_{mj}\to qz_{mt}}$ as a Gaussian distribution. Based on these two approximations, we characterize both $\Delta_{z_{mt}\to qz_{mk}}^{i+1}$ and $\Delta_{ z_{mk}}^{i+1}$ as Gaussian distributions with tractable means and variances.
	\item To further reduce the computational complexity, we show that  
	$\Delta_{ z_{mk}\to qz_{mt}}^{i+1}$ differs from $\Delta_{ z_{mk}}^{i+1}$ in only one term, and this term vanishes in the large-system limit. As a consequence, we use the mean and variance of $\Delta_{ z_{mk}}^{i+1}$ (\ie $\zh_{mk}(i+1)$ and $v^z_{mk}(i+1)$) to approximate those of $\Delta_{ z_{mk}\to qz_{mt}}^{i+1}$, and obtain closed-loop updating formulas for $\zh_{mk}$ and $v^z_{mk}$.   
	\item With the tractable form of $\Delta_{ z_{mk}}^{i+1}$, we show in a similar way that $\Delta_{ w_{ml}}^{i+1}$, $\prod_{m}\Delta_{ zwg_{mk}\to g_{lk}}^i$, and $\prod_{m,l} \Delta_{ws_{ml}\to s_{m^\prime l^\prime}}^i$ can be approximated as Gaussian distributions. Plugging these expressions into \eqref{upost}--\eqref{rpost} and taking the prior information \eqref{gin1}--\eqref{gin2} into account, we obtain the closed-loop updating formulas for $\gh_{lk}$, $v^g_{lk}$, $\sh_{m^\prime l^\prime}$ and $v^s_{m^\prime l^\prime}$.
\end{itemize}
We summarize the resultant algorithm in Algorithm \ref{alg1}. In Algorithm \ref{alg1}, we employ the adaptive damping technique \cite{MP_BIGAMP1} to improve the convergence. The details are omitted due to space limitation.
{{Besides, we terminate Algorithm \ref{alg1} either when the maximum allowable number of iterations $I_{\max}$ is reached or when the change between any two consecutive iterations is smaller than $\e$ for $\e$ being a small positive value.}} 
The simulation code is available at \href{https://github.com/liuhang1994/Matrix-Calibration-Based-Cascaded-Channel-Estimation}{https://github.com/liuhang1994/Matrix-Calibration-Based-Cascaded-Channel-Estimation}.
\begin{algorithm}[!t]\label{A1}
	\caption{The Proposed Algorithm}
	\label{alg1}
	\textbf{Input:} $\Yv; \Av_B; \Hv_0; \Rv; \Xv; \tau_N; \lambda_S; \lambda_G; \tau_S; \tau_{G}.$\\
	\textbf{Initialization}:
	$\gammah_{mt}(0)=\xih_{mk}(0)=\alphah_{ml}(0)=0$;\\ $\sh_{m^\prime l^\prime}(1)$ randomly drawn from $p(s_{m^\prime l^\prime})$;\\
	$\wh_{ml}(1)=h_{0,ml}+\sum_{m^\prime,l^\prime}a_{B,mm^\prime}\sh_{m^\prime l^\prime}(1)r_{l^\prime l}$;\\ $\gh_{lk}(1)=\zh_{mk}(1)=0$; \\$v^s_{m^\prime l^\prime}(1)=v^g_{lk}(1)=v^w_{ml}(1)=v^z_{mk}(1)=1$.\\
	\begin{algorithmic}[0]
		\FOR{$i=1,2,\cdots,I_{\text{max}}$}
		\STATE
		{\noindent
			For $\forall m,k$, update $\zh_{mk}$ and $v^z_{mk}$ by  \eqref{temp52}, \eqref{temp51}, \eqref{pupdate}, \\and  \eqref{vz1}--\eqref{vz}.\\
			For $\forall l,k$, update $\gh_{lk}$ and $v^g_{lk}$ by \eqref{temp57}, \eqref{vb}--\eqref{bh}, \\and \eqref{uupdate}.\\
			For $\forall m,l$, update $\wh_{ml}$ and $v^w_{ml}$ by \eqref{vc}--\eqref{ch}, \eqref{vmu}--\eqref{muh}, and  \eqref{wupdate}.\\
			For $\forall m^\prime,l^\prime$, update $\sh_{m^\prime l^\prime}$ and $v^s_{m^\prime l^\prime}$ by \eqref{temp889}, \eqref{vd}--\eqref{dh},\\ and \eqref{rupdate}.\\
			\IF{$\sqrt{\frac{\sum_{m^\prime}\sum_{l^\prime}\abs{\sh_{m^\prime l^\prime}(i+1)-\sh_{m^\prime l^\prime}(i)}^2 }{\sum_{m^\prime}\sum_{l^\prime}\abs{\sh_{m^\prime l^\prime}(i)}^2 }}\leq \e$ \textbf{and} $\sqrt{\frac{\sum_{l}\sum_{k}\abs{\gh_{lk}(i+1)-\gh_{lk}(i)}^2 }{\sum_{l}\sum_{k}\abs{\gh_{lk}(i)}^2 }}\leq \e$} 
			\STATE{Stop;}   \ENDIF     
		} 
		\ENDFOR\\
	\end{algorithmic}
	\textbf{Output:}{ $\sh_{m^\prime l^\prime}$ and $\gh_{lk}$.
	}
\end{algorithm}
\subsection{Computational Complexity}\label{sec_com}
In this subsection, we analyze the computational complexity of the proposed algorithm. Recall that each iteration of Algorithm \ref{alg1} consists of four steps. The computational complexity of each step is detailed as follows. The complexity of computing $\zh_{mk}$ and $v^z_{mk}$ is $\mathcal{O}(MKT)$; the complexity of computing $\gh_{lk}$ and $v^g_{lk}$ is $\mathcal{O}(MKL^\prime)$; the complexity of computing  $\wh_{ml}$ and $v^w_{ml}$ is $\mathcal{O}(MKL^\prime)$; and the complexity of computing $\sh_{m^\prime l^\prime}$ and $v^s_{m^\prime l^\prime}$ is $\mathcal{O}\left(MM^\prime (L^\prime)^2\right)$. 
As a consequence, the overall complexity of the proposed algorithm is $\mathcal{O}\left(I(MKT+2MKL^\prime+MM^\prime (L^\prime)^2)\right)$, where $I\leq I_{\text{max}}$ is the number of iterations in Algorithm \ref{alg1}.

When the ratios $M/K$, $M^\prime/K$, $L/K$, $L^\prime/K$, and $T/K$ are fixed, the overall complexity of the proposed algorithm can be simplified as $\mathcal{O}\left(IK^4\right)$ with a large $K$. In contrast,
both the MMSE estimators \eqref{temp1000} and the canonical message passing algorithm \eqref{qtog}--\eqref{rpost} involve integrations with respect to $\Sv$ or $\Gv$. Therefore, the corresponding computational complexity grows exponentially with $K^2$.
We conclude that our approximations in Algorithm \ref{alg1} significantly reduce the computational complexity compared with the conventional estimators in \eqref{temp1000} and \eqref{qtog}--\eqref{rpost}.
\section{MSE Analysis by Replica Method}\label{sec_ana}
As discussed in Section \ref{sec_bi}, the MMSEs in \eqref{mseMatrix} are difficult to evaluate in general. 
In this section, we derive an asymptotic performance bound of the MSEs under some mild assumptions by employing the replica method from statistical physics \cite{SpinGlass}. 
Although some basic assumptions used in the replica method cannot be rigorously justified, this method has proved successful in analyzing bilinear matrix factorization problems\cite{Phase_ISIT,MP_BIGAMP3}. Moreover, the result in \cite{MP_BIGAMP3} is adopted in analyzing the symbol error rate of the massive MIMO systems with low-precision analog-to-digital convertors in \cite{MIMO_bayes}. The analysis in this section can be regarded as an extension of the replica framework in \cite{MP_BIGAMP3} to the considered matrix-calibration-based cascaded channel estimation problem in \eqref{eq2}.
\subsection{Asymptotic MSE Analysis}\label{sec_mse}
We assume that the prior distributions \eqref{gout}--\eqref{gin2} are perfectly known. Furthermore, throughout this subsection, we restrict our analysis to the scenario where both the sampling grids $\varphiv$ and $\varsigmav$ uniformly cover $[-1,1]$ with $L_1^\prime=L_1$ and $L_2^\prime=L_2$. As a consequence, the array response matrices $\Av_{B,v}$ and $\Av_{B,h}$ in \eqref{Hrmodel} become two normalized discrete Fourier transform (DFT) matrices and we have $\Rv=\Iv$. Discussions on the case of over-complete sampling bases with $L_1^\prime>L_1$ and/or $L_2^\prime>L_2$ can be found in the next subsection. Under these assumptions, the analysis is conducted by evaluating the average free entropy in the large-system limit, \ie $M,M^\prime,K,L,L^\prime,T,\tau_N\to \infty$ with the ratios $M/K$, $M^\prime/K$, $L/K$, $L^\prime/K$, $T/K$, and $\tau_N/K^2$ fixed. In the sequel, we use $K\to \infty$ to denote this limit for convenience. We define the average free entropy as
\begin{align}\label{FreeE}
	\mathcal{F}\triangleq \lim_{K\to \infty}\frac{1}{K^2}\E_{\Yv}\left[\ln p(\Yv)\right].
\end{align}
As shown in \cite{Phase_ISIT}, the MMSEs in \eqref{mseMatrix} correspond to a stationary point of $\mathcal{F}$. Following the argument in \cite[eq. (35)]{MIMO_bayes}, we first transform \eqref{FreeE} as
  \begin{align}\label{FreeE2}
\mathcal{F}= \lim_{n\to 0}\frac{\partial}{\partial n}\lim_{K\to \infty}\frac{1}{K^2}\ln \E_{\Yv}\left[p^n(\Yv)\right].\end{align}
With \eqref{FreeE2}, we compute $\ln \E_{\Yv}\left[p^n(\Yv)\right]/K^2$ as $K\to \infty$ and then compute the derivative with respect to $n$ with $n\to 0$. Finally, we compute the stationary point of $\mathcal{F}$ to obtain the values of the MMSEs. 

To facilitate the computation, we define 
\begin{subequations}\label{temp120}
	\begin{align}
	&Q_S=\lambda_S\tau_S,Q_G=\lambda_G\tau_G,\\
	&Q_W=\frac{M^\prime }{M}Q_S+\tau_{H_{0}},Q_Z=L^\prime Q_W Q_G,
	\end{align}
\end{subequations}
where $\tau_{H_{0}}\triangleq \E[\abs{h_{0,ml}}^2]$.

Besides, we define two \emph{scalar} AWGN channels as
	\begin{align}\label{Scalar_Channel}
&Y_S=\sqrt{\widetilde m_S}S+N_S,Y_G=\sqrt{\widetilde m_G}G+N_G,
	\end{align}
where $N_S,N_G\sim\CN(\cdot;0,1)$; 
$S\sim p(S)\triangleq(1-\lambda_S)\delta(S)+\lambda_S\CN(S;0,\tau_S )$; 
$G\sim p(G)(1-\lambda_G)\delta(G)+\lambda_G\CN(G;0,\tau_G)$; 
and
the parameters $\widetilde m_S,\widetilde m_G$ will be specified later in \eqref{SE}. 

From Bayes' theorem, we can obtain the posterior distributions of $S$ and $G$
as $p(S|Y_S)\propto p(S)p(Y_S|S)$ and $p(G|Y_G)\propto p(G)p(Y_G|G)$. 
Similarly to \eqref{temp1000}, the corresponding posterior mean estimators are given by $\Sh=\int S p(S|Y_S) \mathrm{d}S$ and $\Gh=\int G p(G|Y_G) \mathrm{d}G$.
Moreover, these estimators achieve the following MSEs:
		\begin{subequations}\label{mse_Scalar}
	\begin{align}
		&\text{MSE}_{S}=\E_{S,Y_S}\!\left[\abs{S-\Sh}^2\right]\!,\\
		&\text{MSE}_{G}=\E_{G,Y_G}\!\left[\abs{G-\Gh}^2\right]\!.
		\end{align}
		\end{subequations}

With these definitions, we show in Appendix \ref{appb} that the MMSEs of $\Sv$ and $\Gv$ in \eqref{mseMatrix} converge to $\text{MSE}_{S}$ and $\text{MSE}_{G}$ as $K\to\infty$, where $(\text{MSE}_{S},\text{MSE}_{G})$ is the non-trivial solution to the following fixed-point equations:
\begin{subequations}\label{SE}
	\begin{align}
&\widetilde m_Z=\frac{T\tau_X}{\tau_N+K\tau_X(Q_Z-m_Z)},\label{SE1}\\
&\widetilde m_W=\frac{Km_G}{1/\widetilde m_Z+Q_Z-L^\prime m_Wm_G},\label{SE2}\\
&\widetilde m_G=\frac{Mm_W}{1/\widetilde m_Z+Q_Z-L^\prime m_Wm_G},\label{SE3}\\
&\widetilde m_S=\frac{1}{1/\widetilde m_W+Q_W-\tau_{H_0}-M^\prime m_S/M}, \label{SE4}\\
&m_Z=Q_Z-\frac{Q_Z-L^\prime m_W m_G}{1+\widetilde m_Z(Q_Z-L^\prime m_W m_G)}\label{SE5}\\
&m_W=Q_W-\frac{Q_W-\tau_{H_0}-M^\prime m_S/M}{1+\widetilde m_W(Q_W-\tau_{H_0}-M^\prime m_S/M)},\label{SE6}\\
& m_G=(1-\lambda_G)\E_{N_G}\left[\Big|{f_G\left(\frac{N_G}{\sqrt{\widetilde m_G}},\frac{1}{\widetilde m_G}\right)}\Big|^2\right]\nonumber\\
&\qquad +\lambda_G\E_{N_G}\left[\Big|{f_G\left(\frac{N_G\sqrt{{\widetilde m_G}+1}}{\sqrt{\widetilde m_G}},\frac{1}{\widetilde m_G}\right)}\Big|^2\right],\label{SE7}\\
& m_S=(1-\lambda_S)\E_{N_S}\left[\Big|{f_S\left(\frac{N_S}{\sqrt{\widetilde m_S}},\frac{1}{\widetilde m_S}\right)}\Big|^2\right]\nonumber\\
&\qquad +\lambda_S\E_{N_S}\left[\Big|{f_S\left(\frac{N_S\sqrt{{\widetilde m_S}+1}}{\sqrt{\widetilde m_S}},\frac{1}{\widetilde m_S}\right)}\Big|^2\right],\label{SE8}\\
& \text{MSE}_{G}=Q_G-m_G,\label{SE9}\\
&\text{MSE}_{S}=Q_S-m_S,\label{SE10}
\end{align}
\end{subequations}
In \eqref{SE7}--\eqref{SE8}, we define
\begin{subequations}
	\begin{align}
		&f_G(x,y)=\frac{\int Gp(G)\CN(G;x,y)\mathrm{d}G}{\int p(G)\CN(G;x,y)\mathrm{d}G},\\
		&f_S(x,y)=\frac{\int Sp(S)\CN(S;x,y)\mathrm{d}S}{\int p(S)\CN(S;x,y)\mathrm{d}S}.
	\end{align}
\end{subequations}

As a result, we can obtain MSEs in \eqref{mse_Scalar} by computing the fixed-point solution of \eqref{SE}, which
asymptotically describes the MMSEs in \eqref{mseMatrix}. This can be
efficiently implemented by iteratively updating $\{\text{MSE}_S,\text{MSE}_G,m_o,\widetilde m_o:o\in\{Z,W,S,G\}\}$ following \eqref{SE} until convergence.
\subsection{Further Discussions}\label{sec_ana2}
It is worth noting that the replica method in Appendix \ref{appb} adopts the CLT to approximate $w_{ml}^{(a)}$, $z_{mk}^{(a)}$, and $q_{mt}^{(a)}$ as Gaussian distributions in \eqref{temp1230}, where $x^{(a)}, 0\leq a \leq n$ is the $a$-th replica of random variable $x$ and follows the same distribution as $x$. To apply the CLT, we restrict the sampling bases $\Av_{B,v}$ and $\Av_{B,h}$ to be two normalized DFT matrices in Section \ref{sec_mse}.
 In a more general case where $\Av_{B,v}$ and $\Av_{B,h}$ are two over-complete bases, the Gaussian approximations in \eqref{temp1230} may be inaccurate and consequently affect the accuracy of the replica method. In this case, 
 $\text{MSE}_{S}$ and $\text{MSE}_{G}$ in \eqref{mse_Scalar} may not exactly correspond to the stationary point of the average free entropy in \eqref{FreeE}. As a consequence, the performance bound derived in Section \ref{sec_mse} may become loose, as verified by the numerical results presented in the next section.
\section{Numerical Results}\label{Simulation}
\subsection{Simulation Results Under Channel Generation Model \eqref{gout}--\eqref{gin2}}
We conduct Monte Carlo simulations to verify the analysis in Section \ref{sec_ana}. In this subsection, we assume that the channel is generated according to the prior distributions in \eqref{gout}--\eqref{gin2}. This allows us to calculate the MMSEs of the cascaded channel estimation problem defined in \eqref{mseMatrix} by using the replica method as in Section \ref{sec_ana}. Note that $\text{MSE}_{S}$ and $\text{MSE}_{G}$ in \eqref{mse_Scalar} derived by the replica method will be used as a benchmark to evaluate the performance of our proposed algorithm (\ie Algorithm \ref{alg1}).\footnote{Although the replica analysis in Section \ref{sec_ana} is valid only in the large-system limit, we still use the derived performance bound as a benchmark even when we conduct simulations with finitely large systems.}

{We first verify the efficiency of Algorithm \ref{alg1} over the conventional MMSE estimators. 
Table \ref{table_com} lists the running time for Algorithm \ref{alg1} and the MMSE estimators \eqref{temp1000} with $M=M^\prime=L=L^\prime=K$, where $I_{\max}$ is set to $2000$ for Algorithm \ref{alg1} and the integrations in computing the MMSE estimators are implemented by numerical quadratures with $2$ integration points per dimension.\footnote{Note that a larger number of integration points will improve the integration accuracy but increase the running time.} The simulation platform is a Dell Optiplex 9010 Desktop with an i$7$--$3770$/$3.40$GHz Quad-Core CPU and $16$GB RAM. All results are obtained by averaging over $100$ Monte Carlo trials. We see that the proposed algorithm runs much faster than the MMSE estimators. This is because the proposed approximations avoid the high-dimensional integrations in directly computing \eqref{temp1000}. In addition, the running time of the MMSE estimators sharply increases as $K$ becomes large. This is consistent with the analysis in Section \ref{sec_com}.
\begin{table}[!t]
\centering
\caption{Running Time in Seconds}\label{table_com}
\medskip
\begin{tabular}{l|l|l}
\toprule
System Size& Algorithm \ref{alg1} & MMSE Estimators \eqref{temp1000}\\
		\hline
		$K=1$&$1.81*10^{-3}$&$6.75*10^{-1}$\\
		\hline
		$K=2$&$2.33*10^{-3}$&$1.02*10^{4}$\\
		\hline
		$K=3$&$2.82*10^{-3}$&$\approx 4.43*10^{10}$\\
		\bottomrule
\end{tabular}
\end{table}
}

\begin{figure}[!t]
	\centering
\includegraphics[width=3 in]{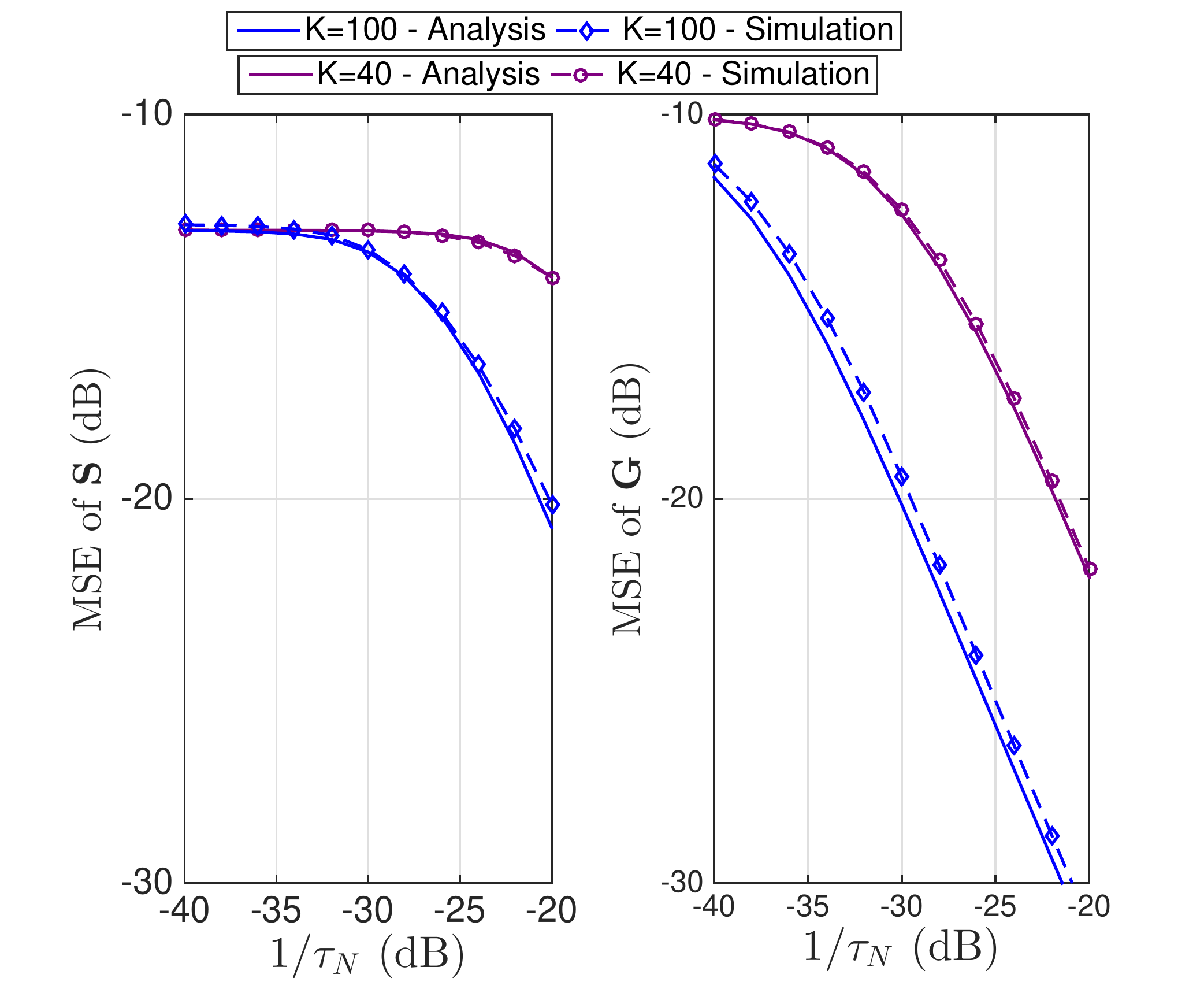}
\caption{MSEs versus $\tau_N$ under normalized DFT bases   $\Av_{B,v}$\protect\\and $\Av_{B,h}$. For $K=40$, we set $L_1=L_1^\prime=4$ and  $L_2=L_2^\prime$\protect\\ $=5$; For $K=100$, we set $L_1=L_1^\prime=$$10$ and $L_2=L_2^\prime=5$.} 
\label{fig_SE}
\end{figure}

Then, we investigate the performance of the proposed algorithm with normalized DFT bases $\Av_{B,v}$ and $\Av_{B,h}$.
We set $M=1.28K$, $M^\prime=1.6K$, $T=1.5K$, $L=L^\prime=0.5K$, $\lambda_G=0.1$, $\lambda_S=0.05$, and $\tau_S=\tau_G=\tau_{H_0}=\tau_{X}=1$.
Fig. \ref{fig_SE} plots the MSE performance versus noise power for $K=40$ and $K=100$. The simulation results are obtained by averaging over $5000$ Monte Carlo trials. We see that the performance of the proposed message passing algorithm closely matches the analytical bound derived by the replica method.   
	\begin{figure}[!t]
	\centering
\includegraphics[width=3in]{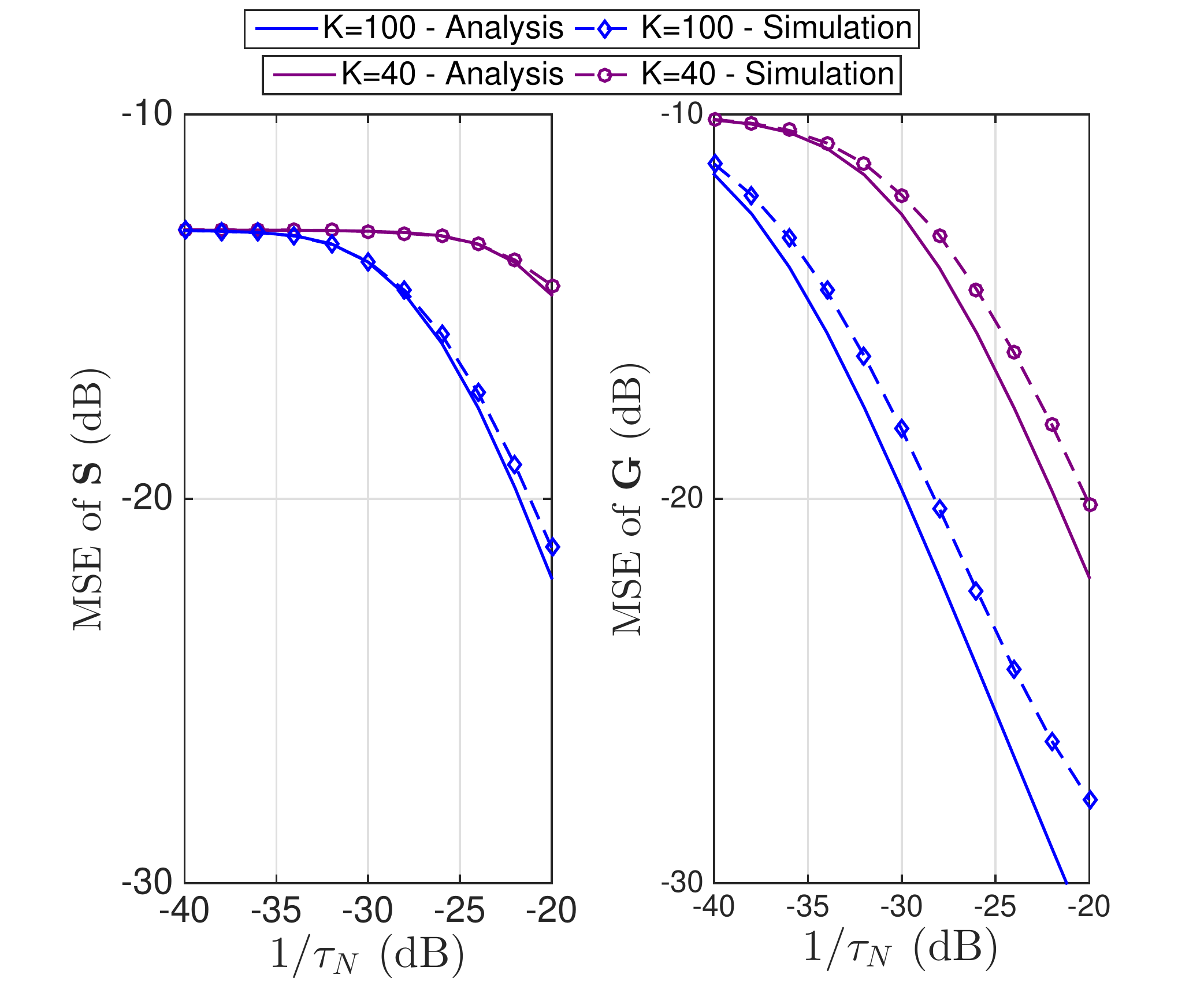}
\caption{MSEs versus $\tau_N$ under over-complete bases $\Av_{B,v}$ and $\Av_{B,h}$. For $K=40$, we set $L_1=L_1^\prime=4$, $L_2=5$, and $L_2^\prime=7$; For $K=100$, we set $L_1=L_1^\prime=10$, $L_2=5$, and $L_2^\prime=7$.}
\label{fig_SE2}
\end{figure}

Next, we simulate on the scenario with an over-complete $\Av_{B,v}$. Specifically, we set $L^\prime=0.7K$ and keep the other parameters unchanged. From Fig. \ref{fig_SE2}, we find that the analytical result has a small gap from the simulation result. As discussed in Section \ref{sec_ana2}, this is because the Gaussian approximations in \eqref{temp1230} are less accurate, and hence the performance bound derived by the replica method is not as tight as in Fig. \ref{fig_SE}.
	\begin{figure}[!t]
	\begin{minipage}[t]{0.5\linewidth}
		\centering
		\includegraphics[width=1.9 in]{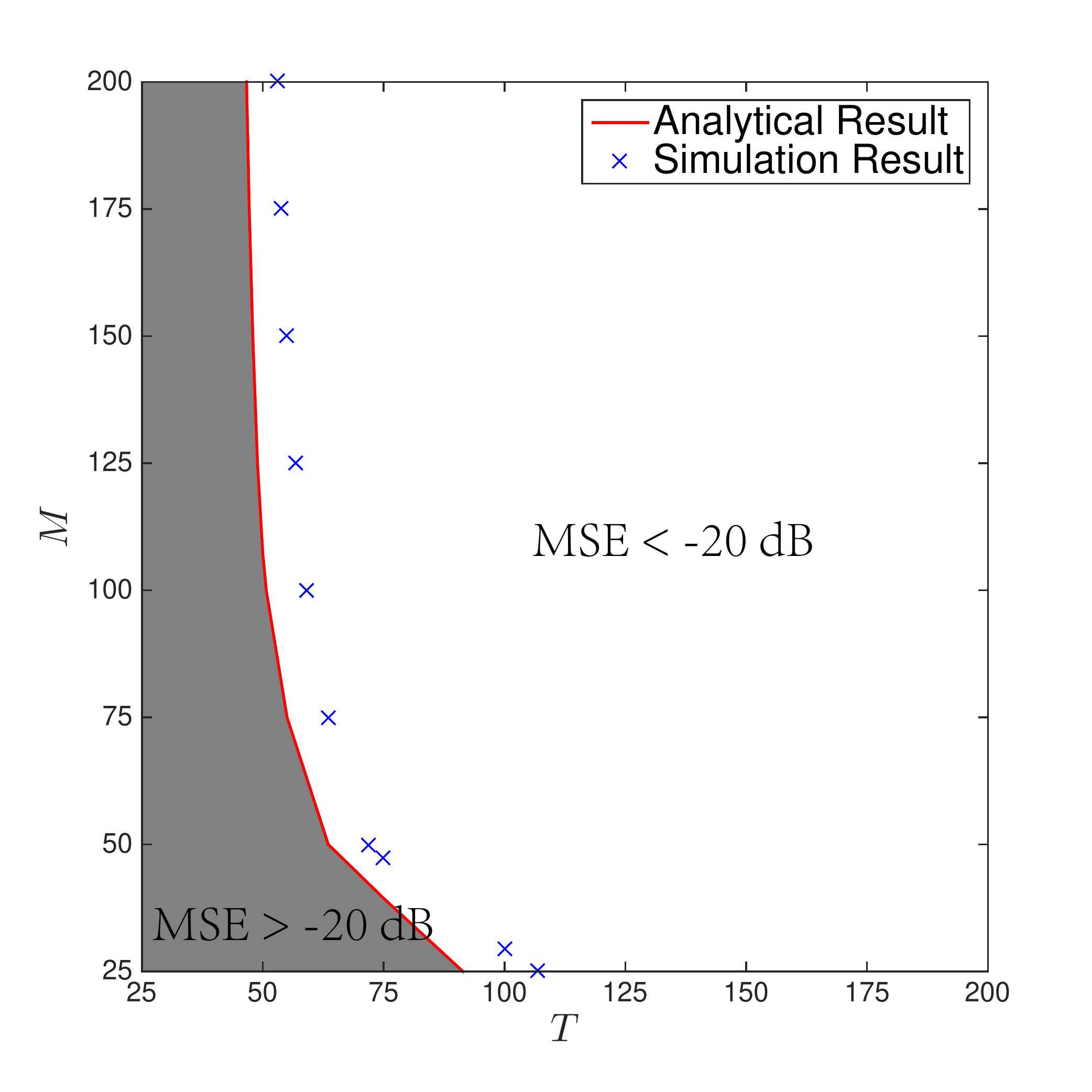}
				\label{fig_pd1}
	\end{minipage}%
	\begin{minipage}[t]{0.5\linewidth}
		\centering
		\includegraphics[width=1.9 in]{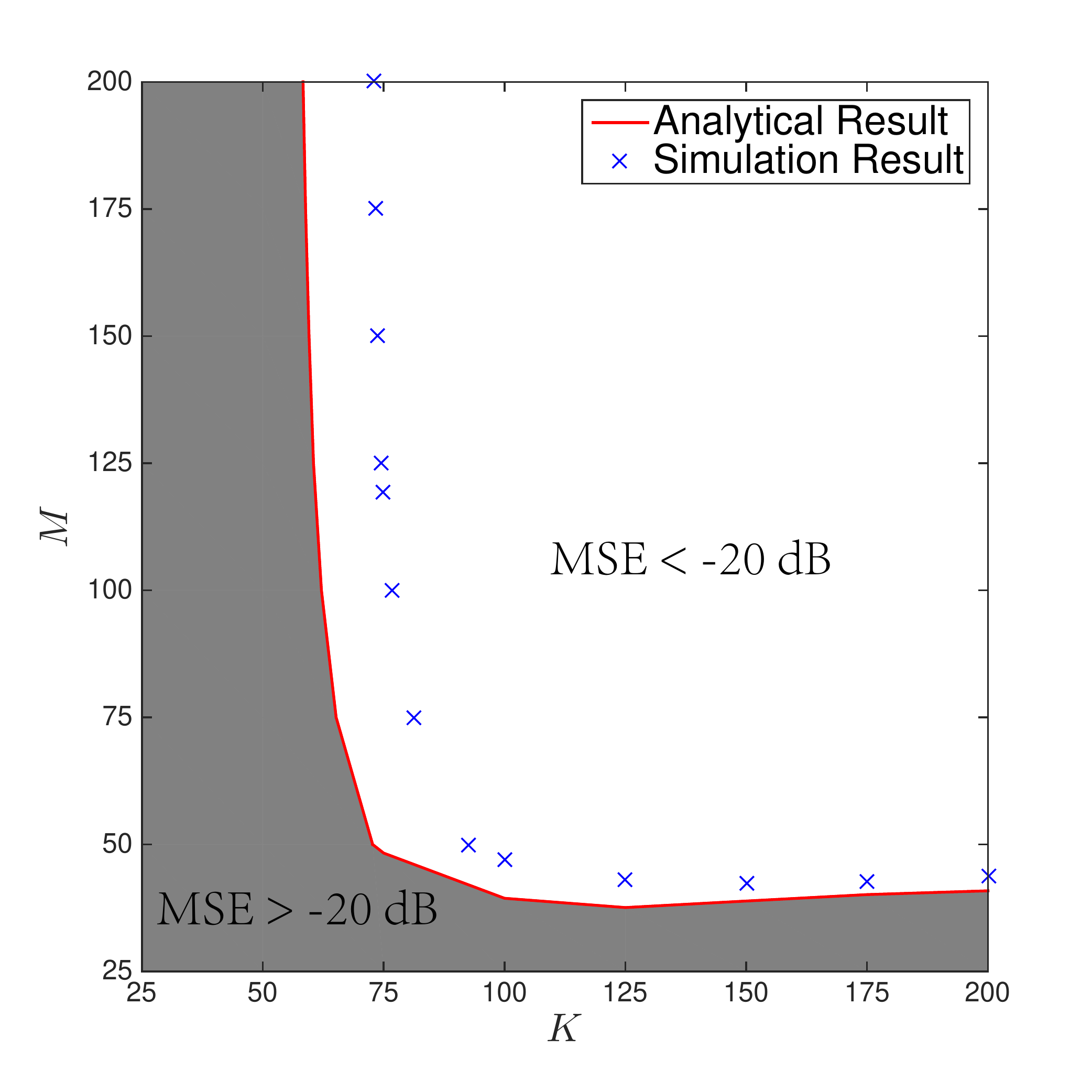}
			\label{fig_pd2}
	\end{minipage}
	\begin{minipage}[t]{0.5\linewidth}
	\centering
	\includegraphics[width=1.9 in]{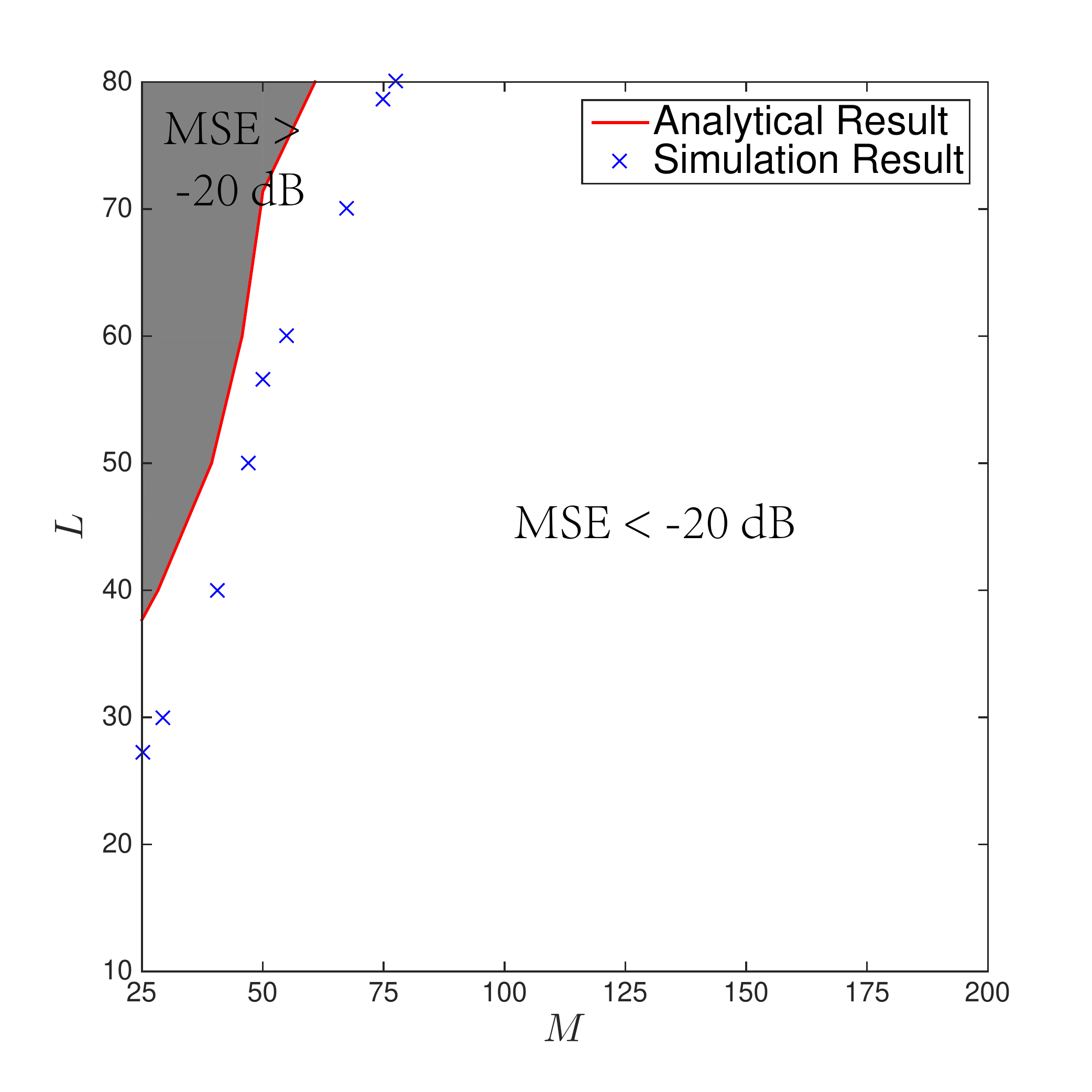}
	\label{fig_pd3}
\end{minipage}%
\begin{minipage}[t]{0.5\linewidth}
	\centering
	\includegraphics[width=1.9 in]{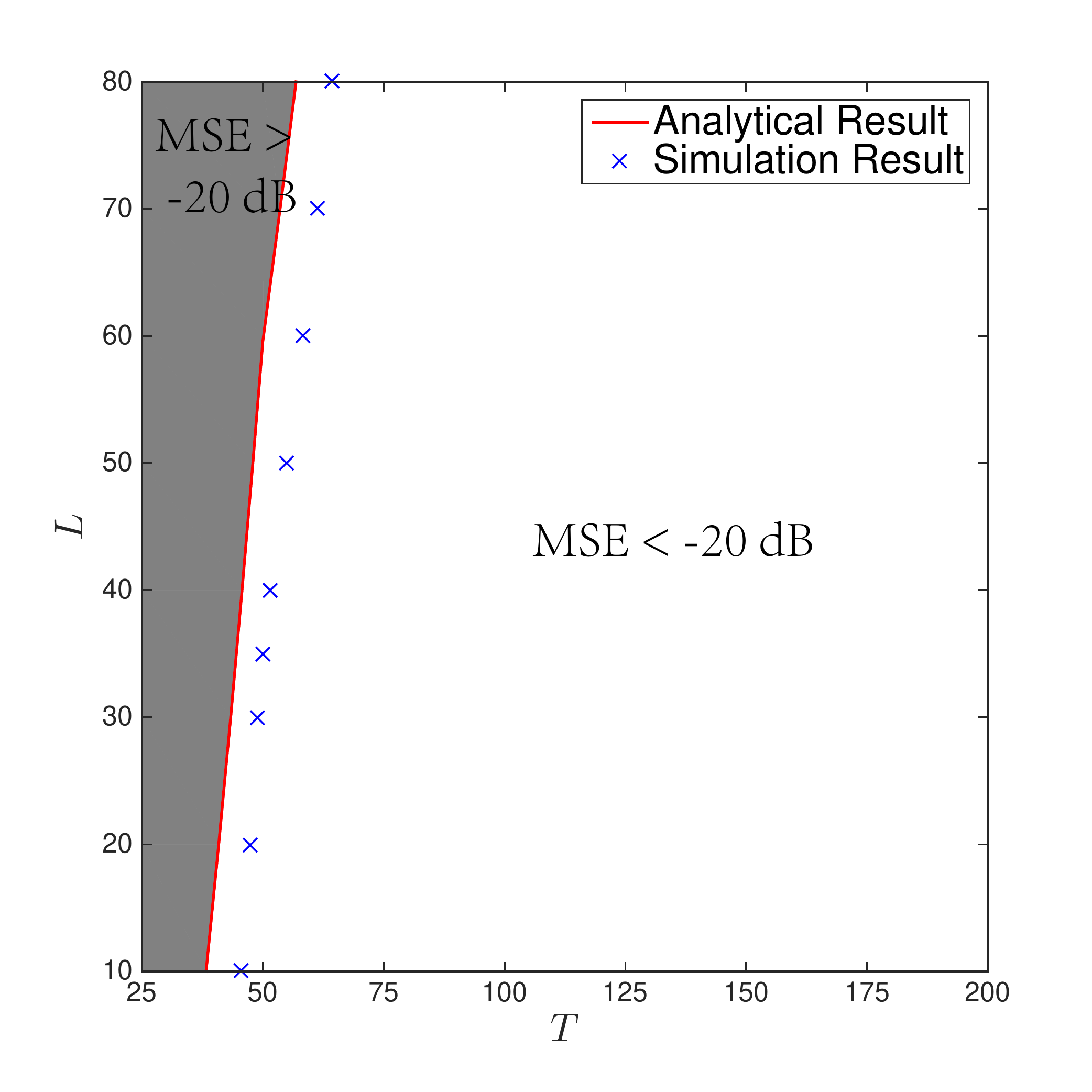}
	\label{fig_pd4}
\end{minipage}
\caption{Phase diagrams with various $M$, $K$, $T$, and $L$. Unless otherwise specified, we set $\lambda_S=0.05$, $\lambda_G=0.1$, $\tau_N=15 \text{dB}$, $K=100$, $M=150$, $M^\prime=1.25M$, $T=75$, $L=50$ and $L^\prime=L$. The red line and the blue cross markers represent the settings that achieve $-20 \text{dB}$ MSEs for the analytical result and Algorithm \ref{alg1}, respectively.}\label{fig_pd}
\end{figure}
%
%

Next, we study the phase transitions of the cascaded channel estimation problem. Here, we say that the estimation of $\Sv$ and $\Gv$ is successful if $\text{MSE}_{\Sv}<-20\text{ dB}$ and $\text{MSE}_{\Gv}<-20\text{ dB}$. 
In Fig. \ref{fig_pd}, we plot the phase transition diagrams for the analytical MSEs \eqref{mse_Scalar} and the empirical MSEs of Algorithm \ref{alg1}.  We find sharp phase transitions for the analytical MSEs, and Algorithm \ref{alg1} requires slightly larger parameters to achieve successful estimation. 
We conclude that the performance of the proposed algorithm is close to the theoretical performance bound in all the simulated settings. 
\begin{table*}[!t]
\caption{Minimum training length $T$ required in different methods}\label{table_overhead}
	\centering
		\begin{tabular}{l|l|l|l|l}
		\toprule
		System Size& Algorithm \ref{alg1}& Replica Analysis & Method in \cite{JSACrevision_3} & Method in \cite{LIS_CE1}\\
				\hline
		$K=100$, $L=50$, $M=128$&$\bf 44$&$11$&$149$&$5000$\\
		\hline
		$K=100$, $L=100$, $M=100$&$\bf 60$&$26$&$199$&$10000$\\
		\bottomrule
	\end{tabular}
\end{table*}

{Finally, we investigate the required training length $T$ for RIS channel estimation in the noiseless case (\ie $\tau_N=0$). In Table \ref{table_overhead}, we list the minimum training length $T$ of the proposed algorithm to achieve near-zero estimation errors ($\text{MSE}_{\Sv}<-50\text{ dB}$ and $\text{MSE}_{\Gv}<-50\text{ dB}$), where we set $\lambda_S=0.05, \lambda_G=0.1$, $M^\prime=1.25M$, and $L^\prime=L$. The performance bound derived from the replica analysis is included as well. Besides, we compare the proposed algorithm with the methods in \cite{JSACrevision_3,LIS_CE1}. Specifically, when $\tau_N =0$, the training length $T$ in \cite{JSACrevision_3} is 
$L+\max\{K-1,\lceil{(K-1)L}/{M} \rceil\}$, where $\lceil\cdot \rceil$ is the ceiling function. And the method in \cite{LIS_CE1} requires $T=LK$. It can be seen that our proposed algorithm requires a smaller $T$ compared with the two baselines. In particular, for the considered settings, the proposed algorithm only needs about $30\%$ of the training overhead of \cite{JSACrevision_3} and about $1\%$ of the overhead of \cite{LIS_CE1}. This is because our algorithm exploits the information on the slowing-varying channel components and the channel sparsity in the angular domain, which significantly reduces the number of variables to be estimated.
}
\subsection{Simulation Results Under a More Realistic Channel Generation Model}
We now consider a more realistic channel generation model as follows.
The large-scale fading component is given by $\beta_i=\beta_{\text{ref}} \cdot d_i^{-\alpha_i},0\leq i\leq K$,
where $\beta_{\text{ref}}$ is the reference path loss at the distance $1$ meter (m); $d_i$ is the corresponding link distance; and $\alpha_i$ is the corresponding pass loss exponent. 
We set $\beta_{\text{ref}}=-20 \text{ dB}$ for all the channel links; $\alpha_0=2$; $\alpha_k=2.6, 1\leq k\leq K$; $d_0=50 \text{ m}$; and $d_k$ uniformly drawn from $[10 \text{ m},12 \text{ m}]$.

Moreover, we generate  $\Hvb_{RB}$ by \eqref{Hrbmodel0} with $20$ clusters of paths and $10$ subpaths per cluster. We draw the central azimuth AoA at the BS of each cluster uniformly over $[-90^\circ,90^\circ]$; draw the central  azimuth (or elevation) AoD at the RIS of each cluster uniformly over $[-180^\circ,180^\circ]$ (or $[-90^\circ,90^\circ]$); and draw each subpath with a $10^\circ$ angular spread. The channels $\Hvt_{RB}$ and $\hv_{UR,k}$ are generated by \eqref{Hrtmodel0} and \eqref{Humodel0} in a similar way both with a cluster of $10$ subpaths. Moreover, every $\alpha_{p}$ is drawn from $\CN(\a_p;0,1)$ and is normalized to satisfy $\norm{\Hv_{RB}}_F^2=\beta_0ML$ and $\norm{\hv_{UR,k}}_2^2=\beta_kL$. We set $K=20$, $M=60$, $T=35$, $L_1=L_2=4$ (\ie $L=16$), $\tau_X=1$, and $\kappa=9$. 
For the proposed algorithm, we set $I_{\text{max}}=2000$, $\e=10^{-4}$,
and the over-complete bases $\varthetav$, $\varphiv$ and $\varsigmav$ in \eqref{Hrmodel}--\eqref{Humodel} to be uniform sampling grids covering $[-1,1]$. The lengths of the sampling grids are set to have a fixed ratio to the antenna dimensions, \ie $M^\prime/M=L_1^\prime/L_1=L_2^\prime/L_2=2$, unless otherwise specified. 
All the results in the sequel are conducted by averaging over $1500$ Monte Carlo trials. 

Apart from the proposed algorithm, the following baselines are involved for comparisons:
\begin{itemize}
	\item Concatenate linear regression (LR): 
	By setting aside $\Hvt_{RB}$, we first set the estimate of $\Hv_{RB}$ as $\hat \Hv_{RB}=\sqrt{{\kappa}/{(\kappa+1)}}  \Hvb_{RB}$, which is assumed to be deterministic.  From \eqref{eq2}, we obtain the following linear regression problem:
	\begin{align}\label{temp121}
	\vect(\Yv)=(\Xv^T\otimes \hat \Hv_{RB}\Av_R)\vect(\Gv)+\nv^\prime,
	\end{align}
	where $\vect(\cdot)$ represents the vectorization operator, and $\nv^\prime$ represents the effective AWGN. 
	We then infer $\Gvh$ from \eqref{temp121} by employing generalized AMP (GAMP) \cite{MP_GAMP2}. Finally, we employ GAMP to estimate $\Hvt_{RB}$ with the estimated $\Gvh$.
	\item Oracle bound with $\Hv_{RB}$ known: Assume that an oracle gives the accurate value of $\Hv_{RB}$. Similarly to \eqref{temp121}, we employ GAMP to obtain $\Gvh$ with $\hat \Hv_{RB}=\Hv_{RB}$. 
	\end{itemize}
	Besides, the following channel estimation methods are included as baselines in the sequel.
	\begin{itemize}
	\item Method in \cite{LIS_CE1}: RIS channels are estimated sequentially. From time slots $(l-1)K+1$ to $lK$, $1\leq l\leq L$,  we turn off all the RIS elements but the $l$-th one. With orthogonal training symbols from the users, the BS computes the LMMSE estimators of the channel coefficients associated with the $l$-th RIS element.
	\item {Method in \cite{JSACrevision_3}: First, the first user sends an all-one training sequence with length no less than $L$ to the BS, and the BS estimates the channel coefficients of this user (\ie $\Hv_{RB}\diag(\hv_{UR,1})$). Then, the other users sequentially send a training symbol to the BS, and the BS estimates $\{\Hv_{RB}\diag(\hv_{UR,k})\}_{2\leq k\leq K}$ by exploiting the correlations with the channel of the first user. }
	\item {Method in \cite{JSACrevision_2}: RIS Channel estimation is divided into $P$ phases with $P=\lfloor T/K\rfloor\leq L$, where $\lfloor \cdot\rfloor$ is the floor function. In the $p$-th phase, the RIS phase-shift vector $\psiv_p$ is set as the $p$-th column of an $L\times L$ DFT matrix. The users send orthogonal training sequences with length no less than $K$ to the BS. The BS collects the received signals in $P$ phases and estimates the cascaded channels $\Hv_{RB}$ and $\{\hv_{UR,k}\}$ alternatively by using parallel factor decomposition. 
	}
	
\end{itemize}
{The estimates of all the three baselines suffer from scaling ambiguities.
\emph{For the comparison purpose}, we remove the scaling ambiguities by assuming the perfect knowledge of $\hv_{UR,1}$. Besides, we set the total training length $T$ to be the same for all the algorithms.
}

We use the normalized MSEs (NMSEs) of $\Hv_{RB}$ and $\{\hv_{UR,k}\}$ to evaluate the performance of the cascaded channel estimation algorithms. Specifically, they are given by
\begin{subequations}
	\begin{align}
	&\text{NMSE of }{\Hv_{RB}}=\frac{\norm{\Hvh_{RB}-\Hv_{RB}}^2_F}{\norm{\Hv_{RB}}^2_F},\\
	&\text{Average NMSE of }{\hv_{UR,k}}=\frac{1}{K}\sum_{k=1}^K\frac{\norm{\hat \hv_{UR,k}-\hv_{UR,k}}^2_2}{\norm{\hv_{UR,k}}^2_2}.
	\end{align}
\end{subequations}
	\begin{figure}[!t]						   
	\centering
	\includegraphics[width=2.8 in]{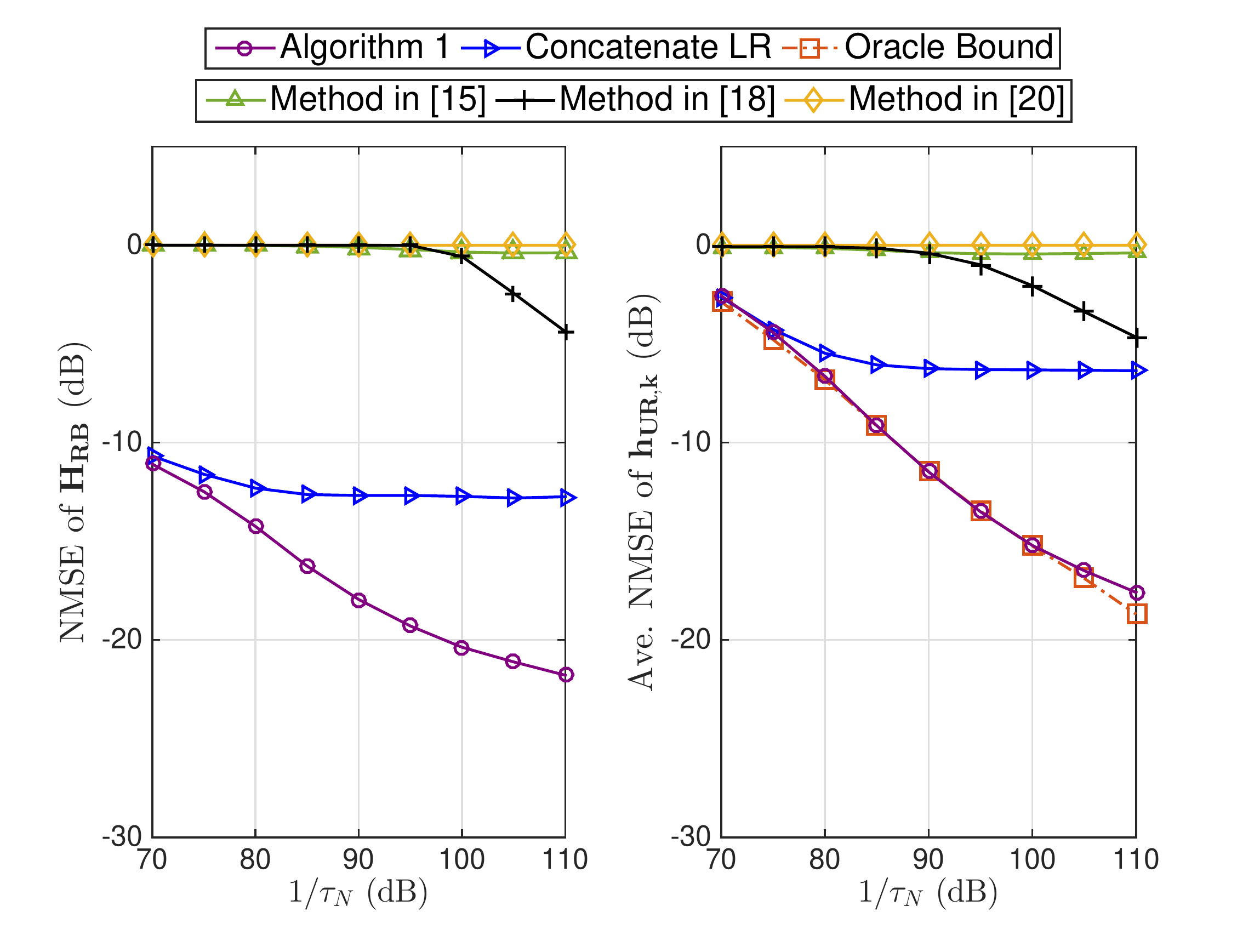}
	\caption{The NMSE performance versus noise power with $T=35$.}
	\label{fig_snr}
\end{figure}

Fig. \ref{fig_snr} investigates the RIS channel estimation performance as $\tau_N$ varies. 
It can be seen that 1) the proposed algorithm achieves an NMSE of $\hv_{UR,k}$ that is very close to the oracle bound, which assumes NMSE of $\Hv_{RB}$ is zero;
2) our algorithm outperforms the baselines, especially when the noise power is large. This is because our proposed algorithm exploits the information on the slowing-varying channel components and the hidden channel sparsity;
 3) the NMSEs of concatenate LR does not decrease when $\tau_N \leq -90 \text{ dB}$. The reason is that the effective noise $\nv^\prime$ in \eqref{temp121} is a combination of  the original AWGN noise $\vect(\Nv)$ and the error resulted from the model mismatch, \ie the ignored term $\left(\Xv^T\otimes\sqrt{1/(\kappa+1)}\Hvt_{RB}\Av_R\right)\vect(\Gv)$. Essentially, the latter term strongly correlates with the variable to be estimated in \eqref{temp121} (\ie $\Gv$). When the model mismatch error dominates the effective noise in the low noise power regime, the correlation issue compromises the convergence of GAMP.  On the contrary, the proposed method avoids this problem since the estimate of $\Hvt_{RB}$, or equivalently $\Sv$, are updated iteratively during the message passing iteration.
	\begin{figure}[!t]
		\centering
			\includegraphics[width=2.8 in]{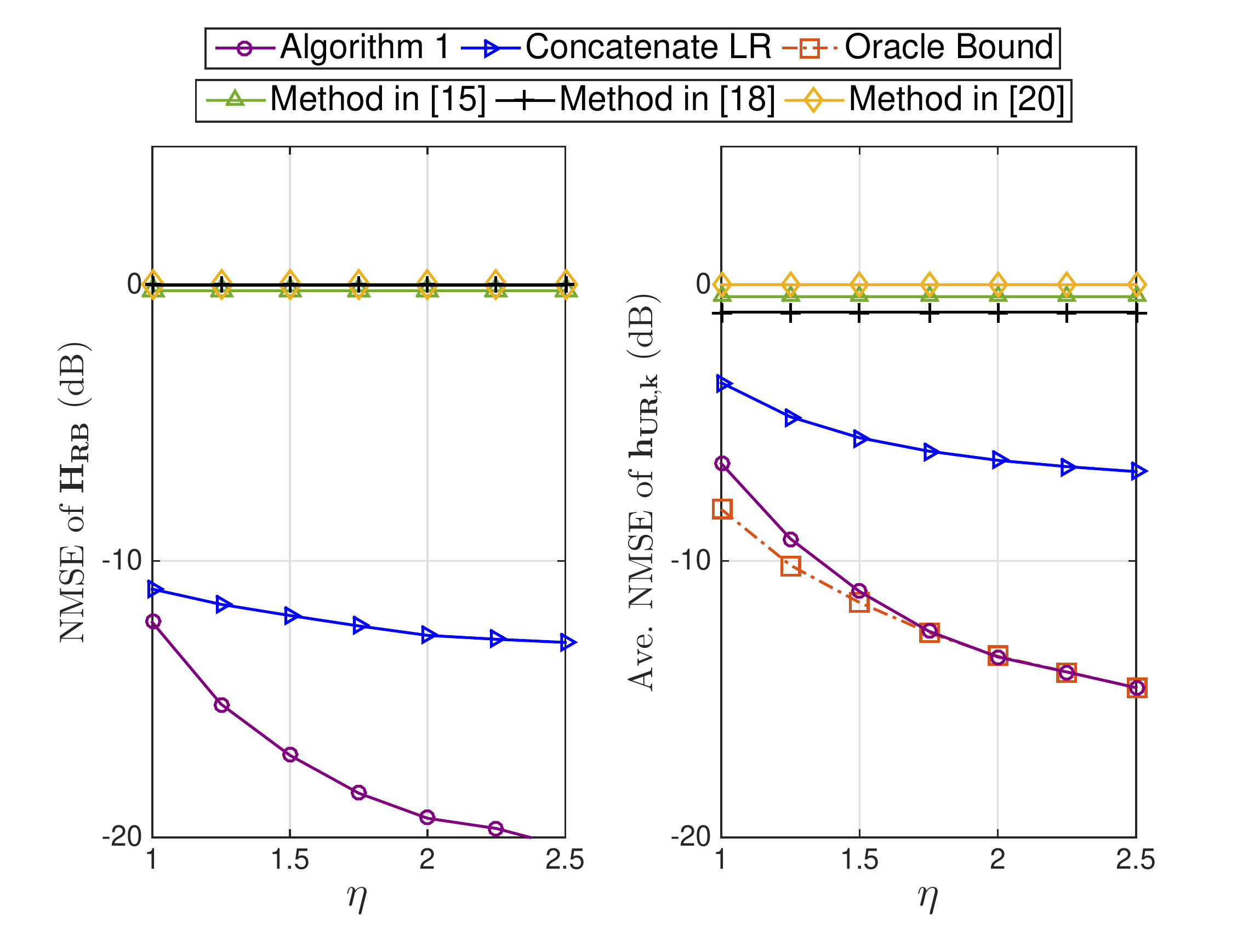}
	\caption{The NMSE performance versus sampling resolution $\eta$, where we set $\tau_N=-95$ dB.}
	\label{fig_eta}
\end{figure}
	\begin{figure}[!t]
		\centering
	\includegraphics[width=2.8in]{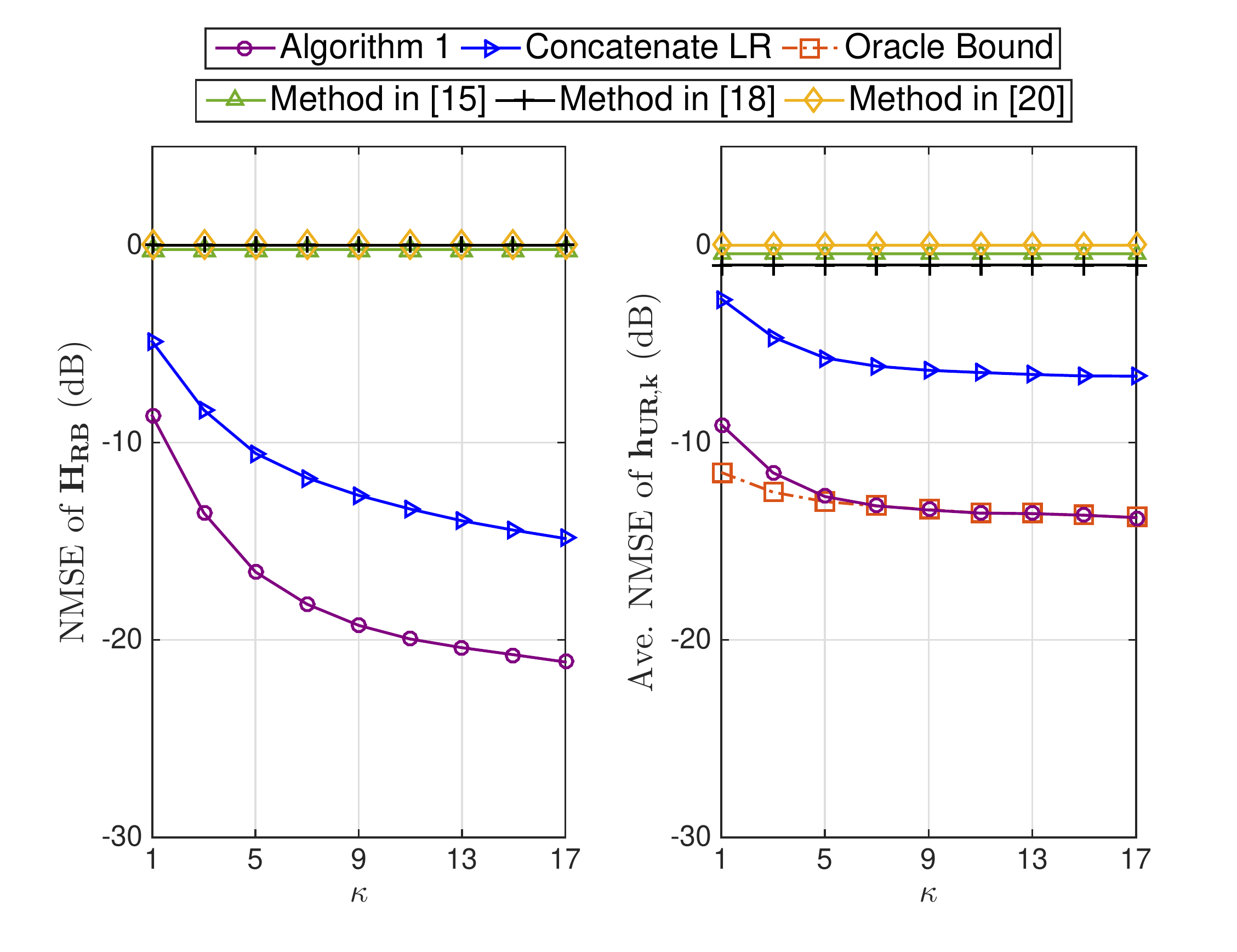}
	\caption{The NMSE performance versus Rician factor $\kappa$ with $\tau_N=-95 \text{ dB}$ and $\eta=2$.}
	\label{fig_kappa}
\end{figure}
\begin{figure}[!t]
		\centering
	\includegraphics[width=2.8in]{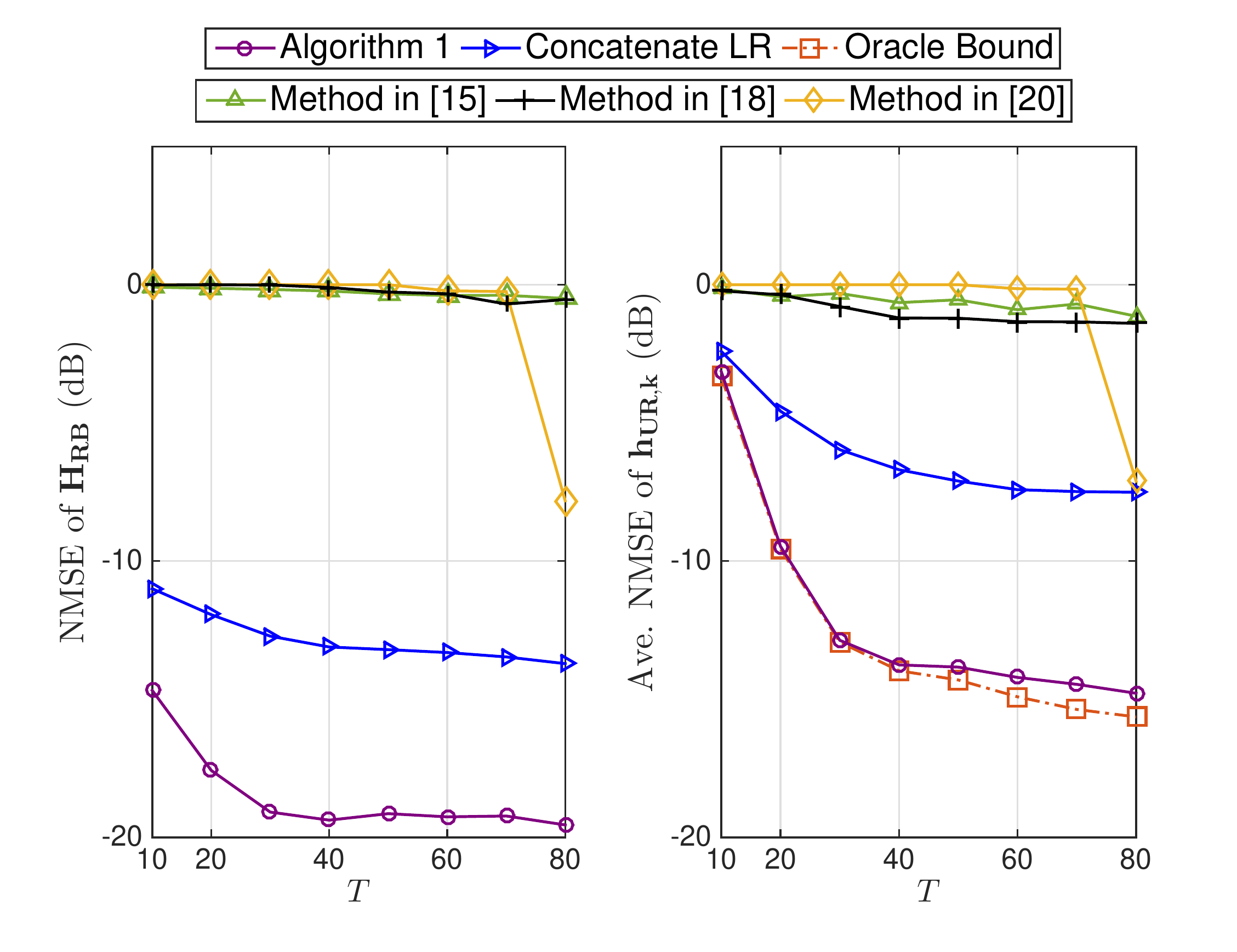}
	\caption{The NMSE performance versus the training length $T$ with $\tau_N=-95 \text{ dB}$.}
	\label{fig_t}
\end{figure}

Next, we study the effect of the grid lengths $M^\prime, L_1^\prime$, and $L_2^\prime$. We use $\eta$ to represent the ratio between the grid length and the number of antennas, \ie $M^\prime/M=L_1^\prime/L_1=L_2^\prime/L_2=\eta$. Fig. \ref{fig_eta} plots the NMSEs of the channel estimation algorithms under various $\eta$, where $\tau_N$ is fixed to $-95 \text{ dB}$.
We have the following observations: 1) the NMSEs of the proposed algorithm, concatenate LR and the oracle estimator decrease as $\eta$ increases, since increasing the sampling grid length leads to a higher angle resolution and hence sparser $\Sv$ and $\Gv$; 2) the baselines \cite{LIS_CE1,JSACrevision_3,JSACrevision_2} do not exploit the channel sparsity in the angular domain and its performance is invariant to $\eta$;
3) the proposed method substantially improves the estimation performance compared to the other algorithms and closely approaches the oracle bound.

In Fig. \ref{fig_kappa}, we study the impact of the Rician factor $\kappa$ on the channel estimation performance. As $\kappa$ increases, the portion of the known information in $\Hv_{RB}$ increases, and hence the NMSEs of the proposed algorithm decrease. Moreover, similarly to Fig. \ref{fig_eta}, we find the proposed method generally achieves better performance compared to the other channel estimation algorithms.

{Finally, Fig. \ref{fig_t} plots the NMSEs versus the training length $T$. We conclude that 1) increasing $T$ from a small value leads to more measurements in $\Yv$, and hence provides sharp improvements to the proposed algorithm;
2) when $T$ is very large, increasing $T$ only provides more redundant measurements and hence slightly improves the estimation performance;
3) the NMSEs for the baselines \cite{LIS_CE1,JSACrevision_3} are only slightly improved with a larger $T$, as they suffer from large estimation errors; 4) the baseline \cite{JSACrevision_2} works with reasonably good performance when $T>70$, which is much larger than the required $T$ of the proposed algorithm ($T\approx 35$).}
\section{Conclusions}\label{sec_conclusion}
In this paper, we studied the channel estimation problem in the RIS-assisted multiuser MIMO system. We formulated the cascaded channel estimation task as a matrix-calibration based sparse matrix factorization problem by exploiting the knowledge of the slow-varying channel components and the hidden channel sparsity in the angular domain. Then, we proposed a novel message-passing based algorithm to infer the cascaded BS-to-RIS and RIS-to-user channels. Furthermore, we developed a framework to analyze the performance of the considered system. Finally, we used numerical results to confirm the performance improvement of the proposed algorithm compared to the state-of-the-art approach.
\appendices
\section{Proof of Proposition \ref{pro1}}
\label{appa0}
Subtracting the term $\sum_{k=1}^K \hv_{UB,k}x_{kt}$ in both sides of \eqref{eq1.1} and letting $\yv(t)\triangleq \yv_0(t)-\sum_{k} \hv_{UB,k}x_{kt}$, we obtain
\begin{align}\label{eq1.2}
\yv(t)&=\sum_{k=1}^K \Hv_{RB}\hv_{UR,k}x_{kt}+\nv(t), 1\leq t \leq T.
\end{align}

With the definitions of $\Xv$, $\Yv$ and $\Nv$, 
we rewrite \eqref{eq1.2} as
\begin{align}\label{eq1}
\Yv&= \Hv_{RB}\Hv_{UR}\Xv+\Nv,
\end{align}
where $\Hv_{UR}=[\hv_{UR,1},\cdots,$ $\hv_{UR,K}]$.
Plugging \eqref{Hrmodel0}, \eqref{Hrmodel}, and \eqref{Humodel} into \eqref{eq1}, we obtain the system model as
\begin{align}\label{eq2app}
\Yv&=\left(\sqrt{\frac{\kappa}{\kappa+1}} \Hvb_{RB}+\Av_{B}\Sv\Av_R^H\right) \Av_R\Gv\Xv+\Nv\nonumber\\
&=\left( \Hv_0+\Av_{B}\Sv\Rv\right) \Gv\Xv+\Nv.
\end{align}

\section{\label{appa}}
\subsubsection{Approximation for \eqref{qtog}--\eqref{pz}} We apply the Fourier inversion theorem to rewrite  \eqref{pz} as
\begin{align}\label{qz2}
\mathcal{P}^i_{z_{mk}}(z_{mk})\propto&\int \mathrm{d}t e^{jtz_{mk}}\prod_{l=1}^{L^\prime}( e^{-jtw_{ml}g_{lk}}\Delta^i_{w_{ml}\to zwg_{mk}}(w_{ml})\nonumber\\
&\times \Delta^i_{g_{lk}\to zwg_{mk}}(g_{lk})\mathrm{d}w_{ml}\mathrm{d}g_{lk} ).
\end{align}
Following the steps in \cite[eqs. (51)--(53)]{MP_BIGAMP3}, we expand the exponential in the bracket in \eqref{qz2} to the second order and simplify \eqref{qz2} as $\mathcal{P}^i_{z_{mk}}(z_{mk})\approx\CN(z_{mk};\ph_{mk}(i),v^p_{mk}(i))$, where
	\begin{subequations}\label{temp145}
	\begin{align}
	v^p_{mk}(i)\!=&\!\sum_{l=1}^{L^\prime}\big(  \abs{\wh_{ml,k}(i)}^2v^g_{lk,m}(i)\!\nonumber\\
	& +\!v^w_{ml,k}(i) \abs{\gh_{lk,m}(i)}^2\!+\!v^w_{ml,k}(i)v^g_{lk,m}(i)\big) ,\\		
	\ph_{mk}(i)\!=&\!\sum_{l=1}^{L^\prime} \wh_{ml,k}(i)\gh_{lk,m}(i). 
	\end{align}
\end{subequations}

Then, we approximate $q_{mt,k}\triangleq \sum_{j\neq k}z_{mj}x_{jt}\sim\prod_{j  \neq k} \Delta^i_{z_{mj}\to qz_{mt}}$ by the CLT as a Gaussian random variable with mean $\qh_{mt}(i)-\zh_{mk,t}(i)x_{kt}$ and variance $v^q_{mt}(i)-v^z_{mk,t}(i)\abs{x_{kt}}^2$, where $\qh_{mt}(i)\triangleq \sum_{k=1}^{K} \zh_{mk,t}(i)x_{kt}$ and $v^q_{mt}(i)\triangleq\sum_{k=1}^{K} v^z_{mk,t}(i)\abs{x_{kt}}^2$. Therefore,
\begin{align}\label{temp140}
&\Delta_{qz_{mt}\to z_{mk}}^i (z_{mk})\propto \int\! \mathrm{d}q_{mt,k}\CN(y_{mt};q_{mt,k}\!+\!z_{mk}x_{kt},\tau_N)\nonumber\\
& \times\CN\left( q_{mt,k};\qh_{mt}(i)\!-\!\zh_{mk,t}(i)x_{kt},v^q_{mt}(i)\!-\!v^z_{mk,t}(i)\abs{x_{kt}}^2\right)\nonumber\\
&\!=\!\CN(y_{mt};\qh_{mt}(i)+(z_{mk}-\zh_{mk,t}(i))x_{kt},\nonumber\\
&\quad \tau_N+v^q_{mt}(i)+(\abs{z_{mk}}^2-v^z_{mk,t}(i))\abs{x_{kt}}^2).
\end{align}
Following the steps in \cite[eqs. (A.6)--(A.16)]{MP_AMPA_arXiv}, we have 
\begin{subequations}
	\begin{align}
&\Delta_{qz_{mt}\to z_{mk}}^{i+1}(z_{mk})= \CN\left(z_{mk};\zh_{mk,t}(i+1),v^z_{mk,t}(i+1)\right),\\ 
&\Delta_{z_{mk}}^{i+1}(z_{mk})= \CN\left(z_{mk};\zh_{mk}(i+1),v^z_{mk}(i+1)\right),\\
&\prod_{t=1}^T\Delta_{qz_{mt}\to z_{mk}}^i (z_{mk})=\CN(z_{mk};\eh_{mk}(i),v^e_{mk}(i)),\label{temp142}
	\end{align}
\end{subequations}
where
	\begin{subequations}\label{temp141}
	\begin{align}
&\zh_{mk,t}(i+1)\approx\zh_{mk}(i+1)-v^z_{mk}(i+1)x_{kt}\gammah_{mt}(i),\\
& v^z_{mk,t}(i+1)\approx v^z_{mk}(i+1),\\
&v^z_{mk}(i+1)=\frac{v^p_{mk}(i)v^e_{mk}(i)}{v^p_{mk}(i)\!+\!v^e_{mk}(i)},\label{vz1}\\ 
&\zh_{mk}(i+1)=\frac{v^p_{mk}(i)\eh_{mk}(i)\!+\!\ph_{mk}(i)v^e_{mk}(i)}{v^p_{mk}(i)\!+\!v^e_{mk}(i)}.\label{vz}
		\end{align}
\end{subequations}
In the above, we introduce the auxiliary variables as
\begin{subequations}\label{temp51}
	\begin{align}
	&v^\gamma_{mt}(i)=\left(v^\beta_{mt}(i)+\tau_N \right)^{-1},\\
	&\gammah_{mt}(i)=v^\gamma_{mt}(i)\left( y_{mt}-\betah_{mt}(i)\right), \label{gammah}\\
	&v^e_{mk}(i)=\left(\sum_{t=1}^T  v^\gamma_{mt}(i)\abs{x_{kt}}^2\right)^{-1},\\
	&\eh_{mk}(i)=\zh_{mk}(i)\!+\!v^e_{mk}(i)\sum_{t=1}^T x_{kt}^\star \gammah_{mt}(i),
	\end{align}
\end{subequations}
where
\begin{subequations}\label{temp52}
	\begin{align}
	&v^\beta_{mt}(i)=\sum_{k=1}^{K}  v_{mk}^z(i)\abs{x_{kt}}^2,\\
	& \betah_{mt}(i)=\sum_{k=1}^{K}  \zh_{mk}(i)x_{kt}\!-\!v^\beta_{mt}(i)\gammah_{mt}(i-1).\label{betah}
	\end{align}
\end{subequations}
\subsubsection{Approximation for \eqref{gtou}--\eqref{wtog} and \eqref{upost}} Plugging  \eqref{temp142} into \eqref{gtou} and \eqref{gtow}, we have
\begin{align}
&\int\mathrm{d} z_{mk}  \prod_{t=1}^T\Delta_{qz_{mt}\to z_{mk}}^i (z_{mk})p(z_{mk}|w_{ml},g_{lk}, \forall l) \nonumber\\
&\!=\!\CN\left( \sum_{l=1}^{L^\prime}w_{ml}g_{lk};\eh_{mk}(i),v^e_{mk}(i)\right) .\label{temp82}
\end{align}
With \eqref{temp82}, the forms of \eqref{gtou} and \eqref{gtow} match eq. (13) of \cite{MP_BIGAMP1}. Following the steps in \cite[Sec. II-D--Sec. II-E]{MP_BIGAMP1}, we obtain 
\begin{subequations}
	\label{temp300}
\begin{align}
&\Delta_{g_{lk}}^{i+1}(g_{lk})\approx p(g_{lk})\CN(g_{lk},\bh_{lk}(i),v^b_{lk}(i))\label{temp143},\\
&\prod_{k=1}^{K} \Delta_{zwg_{mk}\to w_{ml}}^i(w_{ml})=\CN(w_{ml},\ch_{ml}(i),v^c_{ml}(i)).\label{temp144}
\end{align}
\end{subequations}
	
In \eqref{temp300}, we introduce
\begin{subequations}
	\begin{align}
		v^b_{lk}(i)=&\left(\sum_{m=1}^{M} \abs{\wh_{ml}(i)}^2v^\xi_{mk}(i)\right)^{-1},\label{vb}
	\end{align}
	\begin{align}
	\bh_{lk}(i)=& \left(1\!-\!v^b_{lk}(i)\sum_{m=1}^{M} v^w_{ml}(i)v^\xi_{mk}(i)\right)\gh_{lk}(i)\!\nonumber\\
	&+\!v^b_{lk}(i)\sum_{m=1}^{M} \wh_{ml}^{\star}(i)\xih_{mk}(i),\label{bh}\\
	v^c_{ml}(i)=&\left(\sum_{k=1}^{K} \abs{\gh_{lk}(i)}^2v^\xi_{mk}(i)\right)^{-1},\label{vc}\\
	\ch_{ml}(i)=&\left(\!1\!-\!	v^c_{ml}(i)\sum_{k=1}^{K} v^g_{lk}(i)v^\xi_{mk}(i)\!\right)\wh_{ml}(i)\!\nonumber\\
	&+\!	v^c_{ml}(i)\!\sum_{k=1}^{K}\gh_{lk}^{\star}(i)\xih_{mk}(i),\label{ch}
	\end{align}
\end{subequations}
where
\begin{subequations}\label{temp57}
	\begin{align}
v^\xi_{mk}(i)=\frac{v^p_{mk}(i)\!-\!v^z_{mk}(i)}{(v^p_{mk}(i))^2}, \\
	\xih_{mk}(i)=\frac{\zh_{mt}(i)\!-\!\ph_{mt}(i)}{v^p_{mk}(i)}.\label{sh}
	\end{align}
\end{subequations}
Plugging \eqref{gin2} into \eqref{temp143}, we compute the mean and variance of $\Delta_{g_{lk}}^{i+1}$ as	
\begin{subequations}\label{uupdate}
\begin{align}
\gh_{lk}(i+1)&=\int g_{lk}\Delta_{g_{lk}}^{i+1}(g_{lk})\mathrm{d}{g_{lk}},\label{uh}\\
v^g_{lk}(i+1)&=\int g^2_{lk}\Delta_{g_{lk}}^{i+1}(g_{lk})\mathrm{d}{g_{lk}}-\abs{\gh_{lk}(i+1)}^2.\label{vu}
\end{align}
\end{subequations}
Furthermore, following the steps in \cite[Sec. II-F]{MP_BIGAMP1}, we compute \eqref{temp145} as
\begin{subequations}\label{pupdate}
	\begin{align}
	v^p_{mk}(i)=&\sum_{l=1}^{L^\prime}\!\left(  \abs{\wh_{ml}(i)}^2v^g_{lk}(i)\!+\!v^w_{ml}(i) \abs{\gh_{lk}(i)}^2\!+\!v^w_{ml}(i)v^g_{lk}(i)\right)\! ,  \label{vp}\\
	\ph_{mk}(i)=&\sum_{l=1}^{L^\prime} \wh_{ml}(i)\gh_{lk}(i) \!-\!\xih_{mk}(i\!-\!1)\nonumber\\
	&\times\sum_{l=1}^{L^\prime}\left( \abs{\wh_{ml}(i)}^2v^g_{lk}(i)\!+\!v^w_{ml}(i) \abs{\gh_{lk}(i)}^2\right). \label{ph}
	\end{align}
\end{subequations}

\subsubsection{Approximation for \eqref{pw}--\eqref{wpost} and \eqref{rpost}} From \eqref{temp144}, we obtain
\begin{align}
&\int \mathrm{d}w_{ml}\prod_{k=1}^{K} \Delta_{zwg_{mk}\to w_{ml}}^i(w_{ml})p(w_{ml}|s_{m^\prime l^\prime}, \forall m^\prime, l^\prime) \nonumber
\end{align}
\begin{align}
&=\CN\left( \ch_{ml}(i);h_{0,ml}+\sum_{m^\prime=1}^{M^\prime}\sum_{l^\prime=1}^{L^\prime}a_{B,mm^\prime}s_{m^\prime l^\prime}r_{l^\prime l},v^c_{ml}(i)\right) \!.\label{temp87}
\end{align}
Plugging \eqref{temp87} into \eqref{ftor}, we obtain
\begin{align}
&\Delta_{ws_{ml}\to s_{m^\prime l^\prime}}^i(s_{m^\prime l^\prime})\!\propto\!\int\prod_{(j,n)\neq (m^\prime\!,l^\prime)}\left(\Delta_{s_{jn} \to  ws_{ml}}^{i}(s_{jn})\mathrm{d} s_{jn}\right)\nonumber\\
& \times\CN\big( \ch_{ml}(i);h_{0,ml}\!+\!\sum_{m^\prime,l^\prime}a_{B,mm^\prime}s_{m^\prime l^\prime}r_{l^\prime l},v^c_{ml}(i)\big).
\end{align}
Define $\mu_{ml,m^\prime l^\prime}\triangleq \sum_{(j,n)\neq (m^\prime\!,l^\prime)}a_{B,mj}s_{jn}r_{nl}$. By the CLT, we obtain that $\mu_{ml,m^\prime l^\prime}$ is Gaussian distributed with mean $\muh_{ml}(i)-a_{B,mm^\prime}\sh_{m^\prime l^\prime,ml}(i)r_{l^\prime l}$ and variance $v^\mu_{ml}(i)-a_{B,mm^\prime}v^s_{m^\prime l^\prime,ml}(i)r_{l^\prime l}$, where
\begin{subequations}\label{temp301}
	\begin{align}
&	\muh_{ml}(i)=\sum_{m^\prime=1}^{M^\prime}\sum_{l^\prime=1}^{L^\prime}a_{B,mm^\prime}\sh_{m^\prime l^\prime,ml}(i)r_{l^\prime l},\\
&	v^\mu_{ml}(i)=\sum_{m^\prime=1}^{M^\prime}\sum_{l^\prime=1}^{L^\prime} \abs{a_{B,mm^\prime}}^2v^s_{m^\prime l^\prime,ml}(i)\abs{r_{l^\prime l}}^2.
	\end{align}
\end{subequations}
From \eqref{temp301} and following the steps in 
\cite[eqs. (A.6)--(A.16)]{MP_AMPA_arXiv}, we obtain
\begin{subequations}
\begin{align}
\prod_{m^\prime,l^\prime}\Delta^i_{s_{m^\prime l^\prime}\to ws_{ml}}(s_{m^\prime l^\prime})&=\CN\left( \mu_{m l};\muh_{ml}(i),v^\mu_{ml}(i)\right) \!,\label{temp89}\\
\prod_{m^\prime,l^\prime} \Delta_{ws_{ml}\to s_{m^\prime l^\prime}}^i(s_{m^\prime l^\prime})&=\CN\left( s_{m^\prime l^\prime};\dih_{m^\prime l^\prime}(i),v^d_{m^\prime l^\prime}(i)\right)\!,\label{temp90}
\end{align}
\end{subequations}
where
\begin{subequations}\label{temp888}
	\begin{align}
	&v^\mu_{ml}(i)=\sum_{m^\prime=1}^{M^\prime}\sum_{l^\prime=1}^{L^\prime} \abs{a_{B,mm^\prime}}^2v^s_{m^\prime l^\prime}(i)\abs{r_{l^\prime l}}^2,\label{vmu}\\
	&\muh_{ml}(i)=\sum_{m^\prime=1}^{M^\prime}\sum_{l^\prime=1}^{L^\prime}a_{B,mm^\prime}\sh_{m^\prime l^\prime}(i)r_{l^\prime l}-v^\mu_{ml}(i)\alphah_{ml}(i\!-\!1),\label{muh}\\
	&v^d_{m^\prime l^\prime}(i)=\left(\sum_{m=1}^{M}\sum_{l=1}^{L^\prime} \abs{a_{B,mm^\prime}}^2v^\a_{ml}(i)\abs{r_{l^\prime l}}^2\right) ^{-1}\!,\label{vd} \\
	&\dih_{m^\prime l^\prime}(i)=\sh_{m^\prime l^\prime}+v^d_{m^\prime l^\prime}(i)\sum_{m=1}^{M}\sum_{l=1}^{L^\prime}a^\star_{B,mm^\prime}\alphah_{ml}(i)r^\star_{l^\prime l} . \label{dh}
	\end{align}
\end{subequations}
In \eqref{temp888}, we define
\begin{subequations}\label{temp889}
	\begin{align}
	&v^\a_{ml}(i)=\left( {v^\mu_{ml}(i)+v^c_{ml}(i)}\right) ^{-1}, \label{va}\\
	&\alphah_{ml}(i)=v^\a_{ml}(i)\left( \ch_{ml}(i)\!-\!h_{0,ml}\!-\!\muh_{ml}(i)\right) . \label{alphah}
	\end{align}
\end{subequations}

Plugging \eqref{temp89} into \eqref{pw} and combining \eqref{temp144}, we obtain
\begin{subequations}\label{wupdate}
	\begin{align}
	&v^w_{ml}(i+1)=\frac{v^\mu_{ml}(i)v^c_{ml}(i)}{v^\mu_{ml}(i)\!+\!v^c_{ml}(i)},\\
	&\wh_{ml}(i+1)=\frac{v^\mu_{ml}(i)\ch_{ml}(i)\!+\!v^c_{ml}(i)\muh_{ml}(i)\!+\!v^c_{ml}(i)h_{0,ml}}{v^\mu_{ml}(i)\!+\!v^c_{ml}(i)}.\label{wh}
	\end{align}
\end{subequations}

Similarly to \eqref{uupdate}, by plugging \eqref{temp90} into \eqref{rpost}, we obtain $\Delta_{s_{m^\prime l^\prime}}^{i+1}(s_{m^\prime l^\prime})\propto  p(s_{m^\prime l^\prime})\CN\left( s_{m^\prime l^\prime};\dih_{m^\prime l^\prime},v^d_{m^\prime l^\prime}\right)$ and
\begin{subequations}\label{rupdate}
	\begin{align}
	\sh_{m^\prime l^\prime}(i+1)&=\int s_{m^\prime l^\prime}\Delta_{s_{m^\prime l^\prime}}^{i+1}(s_{m^\prime l^\prime})\mathrm{d}{s_{m^\prime l^\prime}},\label{rh} \\
	v^s_{m^\prime l^\prime}(i+1)&=\int s^2_{m^\prime l^\prime}\Delta_{s_{m^\prime l^\prime}}^{i+1}(s_{m^\prime l^\prime})\mathrm{d}{s_{m^\prime l^\prime}}-\abs{\sh_{m^\prime l^\prime}(i+1)}^2.\label{vr}
	\end{align}
\end{subequations}
\section{\label{appb}}
First, note that the replicate partition function $\E_\Yv\left[p^n(\Yv)\right]$ is given by
\begin{align}\label{temp112}
	\E_\Yv\left[p^n(\Yv)\right]=\E_{\boldsymbol{\mathcal{S}},\boldsymbol{\mathcal{G}},\boldsymbol{\mathcal{W}},\boldsymbol{\mathcal{Z}},\boldsymbol{\mathcal{Q}}}\left[\int \mathrm{d}\Yv \prod_{a=0}^n p\left(\Yv|\Qv^{(a)}\right)\right],
\end{align}
where $x^{(a)}, 0\leq a \leq n$ is the $a$-th replica of $x$ and follows the same distribution as $x$; $\boldsymbol{\mathcal{S}}\triangleq\{\Sv^{(a)}:\forall a\}$; $\boldsymbol{\mathcal{G}}\triangleq\{\Gv^{(a)}:\forall a\}$; $\boldsymbol{\mathcal{W}}\triangleq\{\Wv^{(a)}:\forall a\}$; $\boldsymbol{\mathcal{Z}}\triangleq\{\Zv^{(a)}:\forall a\}$; and $\boldsymbol{\mathcal{Q}}\triangleq\{\Qv^{(a)}:\forall a\}$. 

Following \cite{MP_BIGAMP3}, we define the $(n+1)\times(n+1)$ covariance matrix of $\boldsymbol{\mathcal{S}}$ as $\Cv_S\triangleq [C_S^{ab}]_{0\leq a,b,\leq n}$, where $C_S^{ab}=\sum_{m^\prime,l^\prime}(s_{m^\prime l^\prime}^{(a)})^\star s_{m^\prime l^\prime}^{(b)}/(M^\prime L^\prime)$.
Similarly, we define $\Cv_G=[C_G^{ab}]$, $\Cv_W=[C_W^{ab}]=[M^\prime C_S^{ab}/M+\tau_{H_0}]$, and $\Cv_Z=[C_Z^{ab}]=[L^\prime C_W^{ab}C_G^{ab}]$ as the covariance matrices  of $\boldsymbol{\mathcal{G}}$, $\boldsymbol{\mathcal{W}}$, and $\boldsymbol{\mathcal{Z}}$, respectively.

Following \cite[eqs. (219)--(221)]{MP_BIGAMP3}, we obtain
\begin{align}\label{temp122}
	\E_\Yv\left[p^n(\Yv)\right]=&\int\! \E_{\boldsymbol{\mathcal{Q}}|\Cv_Z}\!\left[\int \mathrm{d}\Yv\prod_{a=0}^n p(\Yv|\Qv^{(a)})\right]\mathrm{d}\mu_S(\Cv_S)\nonumber\\
	&\times \mathrm{d}\mu_G(\Cv_G)\mathrm{d}\mu_W(\Cv_W)\mathrm{d}\mu_Z(\Cv_Z),
\end{align}
where
\begin{subequations}
\begin{align}\label{temp113}
	&\mu_S(\Cv_S)\!=\!\E_{\boldsymbol{\mathcal{S}}}\!\left[\prod_{0\leq a\leq b}^n\!\delta\left(\sum_{m^\prime ,l^\prime}\left(s_{m^\prime l^\prime}^{(a)}\right)^\star s_{m^\prime l^\prime}^{(b)}\!-\!M^\prime L^\prime C_S^{ab}\right)\right]\!,\\
	&\mu_G(\Cv_G)\!=\!\E_{\boldsymbol{\mathcal{G}}}\!\left[\prod_{0\leq a\leq b}^n\!\delta\left(\sum_{l,k}\left(g_{lk}^{(a)}\right)^\star g_{lk}^{(b)}\!-\!L^\prime K C_G^{ab}\right)\right]\!,\\
	&\mu_W(\Cv_W)\!=\!\E_{\boldsymbol{\mathcal{W}}}\!\left[\prod_{0\leq a\leq b}^n\!\delta\left(\sum_{m,l}\left(w_{ml}^{(a)}\right)^\star w_{ml}^{(b)}\!-\!M L^\prime C_W^{ab}\right)\right]\!,\\
	&\mu_Z(\Cv_Z)\!=\!\E_{\boldsymbol{\mathcal{Z}}}\!\left[\prod_{0\leq a\leq b}^n\!\delta\left(\sum_{m,k}\left(z_{mk}^{(a)}\right)^\star z_{mk}^{(b)}\!-\!M K C_Z^{ab}\right)\right]\!.
\end{align}
\end{subequations}

By introducing auxiliary matrices $\widetilde\Cv_o=[\widetilde C_o^{ab}], o\in\{S,G,W,Z\}$ and applying the saddle point method \cite[eq. (223)]{MP_BIGAMP3}, we obtain 
\begin{align}\label{temp114}
	\frac{1}{K^2}\ln\E_\Yv\left[p^n(\Yv)\right]=\extr_{\Cv_o,\widetilde\Cv_o,o\in\{S,G,W,Z\}}\mathcal{F}_n,
\end{align}
with
\begin{align}
\mathcal{F}_n\triangleq	\frac{1}{K^2}(\mathcal{I}_Q+\mathcal{I}_Z+\mathcal{I}_W+\mathcal{I}_G+\mathcal{I}_S),
\end{align}
where $\extr$ is the operation of extremization,
\begin{subequations}\label{temp1150}
\begin{align}
	&\mathcal{I}_Q=MT\cdot\ln\E_{\{q^{(a)}\}|\Cv_Z}\!\left[\int \mathrm{d}y\prod_{a=0}^n\CN\left(y;q^{(a)},\tau_N\right)\right],\label{temp115}\\
	&\mathcal{I}_Z=\ln\E_{\boldsymbol{\mathcal{Z}}}\left[e^{\tr(\widetilde\Cv_Z(\Zv^\prime)^H\Zv^\prime)}\right]-MK\tr(\widetilde\Cv_Z\Cv_Z),
	\end{align}
	\begin{align}
	&\mathcal{I}_W=\ln\E_{\boldsymbol{\mathcal{W}}}\left[e^{\tr(\widetilde\Cv_W(\Wv^\prime)^H\Wv^\prime)}\right]-ML^\prime\tr(\widetilde\Cv_W\Cv_W),\\
	&\mathcal{I}_G=\ln\E_{\boldsymbol{\mathcal{G}}}\left[e^{\tr(\widetilde\Cv_G(\Gv^\prime)^H\Gv^\prime)}\right]-L^\prime K\tr(\widetilde\Cv_G\Cv_G),\\
	&\mathcal{I}_S=\ln\E_{\boldsymbol{\mathcal{S}}}\left[e^{\tr(\widetilde\Cv_S(\Sv^\prime)^H\Sv^\prime)}\right]-M^\prime L^\prime\tr(\widetilde\Cv_S\Cv_S).\label{temp118}
\end{align}
\end{subequations}
In \eqref{temp1150}, we define $\Zv^\prime=[z_{mk}^{(a)}]\in\Complex^{MK\times(n+1)}$, $\Wv^\prime=[w_{ml}^{(a)}]\in\Complex^{ML^\prime\times(n+1)}$, $\Gv^\prime=[g_{lk}^{(a)}]\in\Complex^{L^\prime K\times(n+1)}$, and $\Sv^\prime=[s_{m^\prime l^\prime}^{(a)}]\in\Complex^{M^\prime L^\prime\times(n+1) }$ as the matrix representations of $\boldsymbol{\mathcal{Z}}$, $\boldsymbol{\mathcal{W}}$, $\boldsymbol{\mathcal{G}}$, and $\boldsymbol{\mathcal{S}}$, respectively.

Plugging \eqref{temp114} into \eqref{FreeE2}, we have
\begin{align}\label{temp303}
	\mathcal{F}=\lim_{n\to 0}\frac{\partial}{\partial n}\extr_{\Cv_o,\widetilde\Cv_o}\mathcal{F}_n.
\end{align}
To obtain an explicit expression of \eqref{temp303}, we further impose the restrictions on $\Cv_o,\widetilde\Cv_o,o\in\{S,G,W,Z\}$ following the replica symmetry ansatz \cite{MP_BIGAMP3}, as $\Cv_o=(Q_o+m_o)\Iv-m_o {\bf 11}^T$ and $\widetilde \Cv_o=(\widetilde Q_o+\widetilde m_o)\Iv-\widetilde m_o {\bf 11}^T$. Plugging $\Cv_o$ and $\widetilde\Cv_o$ into \eqref{temp303}, we obtain
 \begin{align}\label{temp304}
	\mathcal{F}=\lim_{n\to 0}\frac{\partial}{\partial n}\extr_{\{Q_o,m_o,\widetilde Q_o, \widetilde m_o\}}\mathcal{F}_n.
\end{align}

Furthermore, we approximate $w_{ml}^{(a)}=h_{0,ml}+ \sum_{m^\prime,l^\prime}$ $a_{B,mm^\prime}s_{m^\prime l^\prime}^{(a)}r_{l^\prime l}$, $z_{mk}^{(a)}=\sum_{l}w_{ml}^{(a)}g_{lk}^{(a)}$, and $q_{mt}^{(a)}=\sum_{k}z_{mk}^{(a)}x_{kt}$ as Gaussian random variables by applying the CLT under the large-system limit. Specifically, we have
\begin{subequations}\label{temp1230}
\begin{align}
	w_{ml}^{(a)}&=\sqrt{Q_W-\tau_{H_0}-\frac{M^\prime m_S}{M}}u_W^{(a)} +\sqrt{\tau_{H_0}+\frac{M^\prime m_S}{M}}v_W,\label{temp123}
	\end{align}
	\begin{align}
	z_{mk}^{(a)}&=\sqrt{Q_Z-L^\prime m_Wm_G}u_Z^{(a)}+\sqrt{L^\prime m_Wm_G}v_Z,\label{temp1235}\\
	q_{mt}^{(a)}&=\sqrt{K\tau_X(Q_Z-m_Z)}u_Q^{(a)}+\sqrt{K\tau_X m_Z}v_Q,\label{temp124}
	\end{align}
	\end{subequations}
where $u_W^{(a)},u_Z^{(a)},u_Q^{(a)},v_W,v_Z,v_Q\sim\CN(\cdot;0,1)$.

Substituting \eqref{temp1230} into \eqref{temp304} and following \cite[eqs. (75)--(80)]{MIMO_bayes}, we compute the explicit expression of $\mathcal{F}_n$. By noting $\lim_{n\to 0}\E_\Yv[p^n(\Yv)]=1$, we obtain \eqref{temp120} and $\widetilde Q_o=0$. Moreover, setting the derivative of $\mathcal{F}_n$ with respect to $m_o$ and $\widetilde m_o$􏰕, we obtain \eqref{SE1}--\eqref{SE6}, $m_G=\E_{G,Y_G}[\abs{\hat G}^2]$ and $m_S=\E_{S,Y_S}[\abs{\hat S}^2]$, where $\hat S$ and $\hat G$ are defined in \eqref{mse_Scalar}.
Substituting $p(G)$, $p(S)$, $p(Y_G|G)$, and $p(S|Y_S)$ from \eqref{Scalar_Channel}, we obtain \eqref{SE7} and \eqref{SE8}.
Finally, aligned with the argument in \cite[eq. (241)]{MP_BIGAMP3}, we have \eqref{SE9} and \eqref{SE10}.

As a result, taking the partial derivative of $\mathcal{F}_n$ with respect to $n$ and letting $n \to 0$, we obtain
\begin{align}\label{temp117}
\mathcal{F}=&\frac{2MT}{K^2}\mathcal{J}_Q+\frac{M}{K}(\mathcal{J}_Z-\widetilde m_Z m_Z)+\frac{ML^\prime}{K^2}(\mathcal{J}_W-\widetilde m_W m_W)\nonumber\\
&+\frac{L^\prime}{K}(\mathcal{J}_G-\widetilde m_G m_G)+\frac{M^\prime L^\prime}{K^2}(\mathcal{J}_S-\widetilde m_S m_S),
\end{align}
where
\begin{subequations}
\begin{align}
	&\mathcal{J}_Q=\ln \E_{v,u^{\prime},y}\!\Big[\ln\E_{u}\!\left[\Norm\!\left(y;\!\sqrt{c_1}u\!+\!\sqrt{c_2}v,\tau_N\right)\right]\!\Big]\!,
\end{align}
with $v,u^{\prime},u\sim\Norm(\cdot;0,1)$; $c_1\triangleq K\tau_X(Q_Z-m_Z)$; $c_2\triangleq K\tau_X m_Z$;
and $y\sim\Norm(y;\sqrt{c_1} u^{\prime}+\sqrt{c_2}v,\tau_N)$;
\begin{align}
&\mathcal{J}_Z=\E_{Y_Z}\!\left[\ln\E_{Z^{(1)}}\!\left[e^{-\widetilde m_Z\abs{Z^{(1)}}^2+2\sqrt{\widetilde m_Z}\Re(Y_Z^\star Z^{(1)})}\right]\right]\!,
\end{align}
with $Y_Z\sim \CN(Y_Z;\sqrt{\widetilde m_Z}Z^{(0)},1)$ 
and $Z^{(0)},Z^{(1)}$ defined as in \eqref{temp1235};
\begin{align}
&\mathcal{J}_W=\E_{Y_W}\!\left[\ln\E_{W^{(1)}}\!\left[e^{-\widetilde m_W\abs{W^{(1)}}^2+2\sqrt{\widetilde m_W}\Re(Y_W^\star W^{(1)})}\right]\right]\!,
\end{align}
with $Y_W\sim \CN(Y_W;\sqrt{\widetilde m_W}W^{(0)},1)$ 
and $W^{(0)},W^{(1)}$ defined as in \eqref{temp123};
\begin{align}
&\mathcal{J}_G=\E_{Y_G}\!\left[\ln\E_{G^{(1)}}\!\left[e^{-\widetilde m_G\abs{G^{(1)}}^2+2\sqrt{\widetilde m_G}\Re(Y_G^\star G^{(1)})}\right]\right]\!,\\
&\mathcal{J}_S=\E_{Y_S}\!\left[\ln\E_{S^{(1)}}\!\left[e^{-\widetilde m_S\abs{S^{(1)}}^2+2\sqrt{\widetilde m_S}\Re(Y_S^\star S^{(1)})}\right]\right]\!,
\end{align}
with $Y_G,Y_S$ defined in \eqref{Scalar_Channel}; $G^{(1)} \sim p(G)$; and $S^{(1)} \sim p(S)$.
\end{subequations} 

Finally, by evaluating the stationary points of  \eqref{temp117}, we obtain the MSEs defined in \eqref{mse_Scalar} as the asymptotic MMSEs.
\ifCLASSOPTIONcaptionsoff
\newpage
\fi
\ifhavebib
{
	\bibliographystyle{IEEEtran}

}
\else{
}
\fi
\end{document}